\documentclass[journal=jacsat,manuscript=article,layout=twocolumn]{achemso}
\usepackage[fontsize=10pt]{fontsize}

\usepackage[version=3]{mhchem} 
\usepackage{multirow} 
\usepackage{hyperref}
\hypersetup{
    hidelinks,           
    colorlinks=true,     
    linkcolor=black,     
    urlcolor=blue,       
    citecolor=black,     
    filecolor=black      
}
\usepackage{dirtytalk}
\usepackage[dvipsnames,table]{xcolor}
\usepackage{lmodern}
\usepackage{makecell}
\usepackage{booktabs}  
\usepackage{array}     
\definecolor{highlight}{RGB}{242,242,242}  
\cellcolor{highlight}
\usepackage{subcaption}
\definecolor{lightyellow}{RGB}{255,255,224}
\definecolor{lightblue}{RGB}{240,248,255}
\definecolor{lightgray}{RGB}{242,242,242}
\SectionNumbersOn 

\author{Aravindh Nivas Marimuthu}
\email{nivasm@mit.edu}
\affiliation[MIT]
{Department of Chemistry, Massachusetts Institute of Technology, Cambridge, MA 02139, USA.}
\author{Brett A. McGuire}
\email{brettmc@mit.edu}
\affiliation[MIT]
{Department of Chemistry, Massachusetts Institute of Technology, Cambridge, MA 02139, USA.}
\alsoaffiliation[NRAO]
{National Radio Astronomy Observatory, Charlottesville, VA 22903, USA.}

\title[title]
  {A Machine Learning Pipeline for Molecular Property Prediction using ChemXploreML}


\begin{document}

\begin{tocentry}




\centering
\includegraphics[scale=0.20,keepaspectratio]{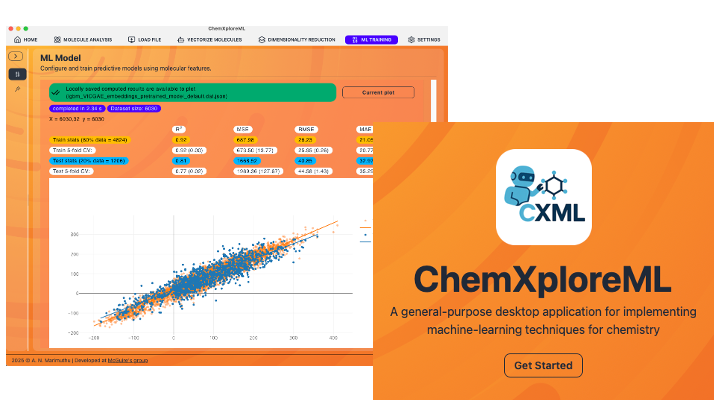}

\end{tocentry}

\begin{abstract}

We present ChemXploreML, a modular desktop application designed for machine learning-based molecular property prediction. The framework's flexible architecture allows integration of any molecular embedding technique with modern machine learning algorithms, enabling researchers to customize their prediction pipelines without extensive programming expertise. To demonstrate the framework's capabilities, we implement and evaluate two molecular embedding approaches - Mol2Vec and VICGAE (Variance-Invariance-Covariance regularized GRU Auto-Encoder) - combined with state-of-the-art tree-based ensemble methods (Gradient Boosting Regression, XGBoost, CatBoost, and LightGBM). Using five fundamental molecular properties as test cases - melting point (MP), boiling point (BP), vapor pressure (VP), critical temperature (CT), and critical pressure (CP) - we validate our framework on a dataset from the CRC Handbook of Chemistry and Physics. The models achieve excellent performance for well-distributed properties, with R$^2$ values up to 0.93 for critical temperature predictions. Notably, while Mol2Vec embeddings (300 dimensions) delivered slightly higher accuracy, VICGAE embeddings (32 dimensions) exhibited comparable performance yet offered significantly improved computational efficiency. ChemXploreML's modular design facilitates easy integration of new embedding techniques and machine learning algorithms, providing a flexible platform for customized property prediction tasks. The application automates chemical data preprocessing (including UMAP-based exploration of molecular space), model optimization, and performance analysis through an intuitive interface, making sophisticated machine learning techniques accessible while maintaining extensibility for advanced cheminformatics users.


\end{abstract}
\section{Introduction}

The prediction of molecular properties plays a crucial role in the chemical sciences, enabling the rapid screening of compounds and accelerating the discovery of materials with desired characteristics \cite{david_molecular_2020, janet_machine_2020, kulik_making_2020}. Properties such as the melting point (MP), boiling point (BP), vapor pressure (VP), critical temperature (CT), and critical pressure (CP) are fundamental to understanding molecular behavior and are essential for both industrial applications \cite{korsten_internally_2000, breedveld_thermodynamic_1973} and theoretical studies in chemistry \cite{marano_general_1997, lucas_prediction_2007, whiteside_estimating_2016, dearden_quantitative_2003}. Traditional experimental methods for determining these properties are time-consuming and resource-intensive \cite{hay_clinical_2014, dowden_trends_2019}, making computational prediction approaches increasingly valuable in modern chemical research.

Recent advances in machine learning (ML) have revolutionized the way we approach molecular property prediction. The key challenge in applying ML to chemical problems lies in transforming molecular structures into machine-readable numerical representations while preserving essential chemical information. Various molecular embedding techniques have emerged to address this challenge, with methods like Mol2Vec \cite{jaeger_mol2vec_2018} and VICGAE (Variance-Invariance-Covariance regularized GRU Auto Encoder) \cite{lee_language_2021} showing particular promise. These approaches aim to capture both critical structural and chemical features of molecules in high-dimensional vector spaces, thereby enabling more accurate property predictions \cite{scolati_explaining_2023, lee_machine_2021}.

The effectiveness of molecular property prediction depends not only on the quality of molecular representations but also on the choice of regression models and careful data preprocessing. Tree-based ensemble methods, including Gradient Boosting Regression (GBR) \cite{friedman_greedy_2001}, XGBoost \cite{chen_xgboost_2016}, LightGBM (LGBM)\cite{ke_lightgbm_2017}, and CatBoost \cite{prokhorenkova_catboost_2018} have demonstrated remarkable success in handling complex structure-property relationships. These models can capture non-linear relationships and handle the high dimensionality inherent in molecular data while maintaining interpretability \cite{boldini_practical_2023}.

To facilitate the application of these ML techniques in chemical research, we have developed ChemXploreML, a comprehensive desktop application designed for molecular property prediction \cite{aravindh_nivas_marimuthu_aravindhnivaschemxploreml_2025}. This application integrates data analysis, preprocessing, and machine-learning modeling into a unified workflow. ChemXploreML accepts various data formats and provides automated analysis of molecular property distributions and structural characteristics through SMILES (Simplified Molecular Input Line Entry System) \cite{weininger_smiles_1988, oboyle_towards_2012} and SELFIES (\textbf{SELF}-referenc\textbf{I}ng \textbf{E}mbedded \textbf{S}trings) \cite{krenn_self-referencing_2020} string parsing. The application performs a systematic examination of the atomic, structural, and elemental distributions, crucial to understanding the characteristics of the dataset and potential biases.

Our study focuses on evaluating the performance of different molecular embedding techniques combined with modern regression algorithms for predicting key molecular properties. We analyze the effectiveness of Mol2Vec and VICGAE embeddings across multiple properties, comparing their ability to capture relevant chemical information and their impact on prediction accuracy. Through this analysis, we aim to provide insights into the optimal combination of embedding methods and regression models for specific molecular property prediction tasks. This comprehensive evaluation is particularly relevant as the field moves towards more sophisticated ML applications in chemistry. Understanding the strengths and limitations of different approaches is crucial for developing more effective prediction models, which, in turn, accelerates the discovery and development of new chemical compounds with desired properties.

In the next section, we present the ChemXploreML desktop application, which integrates these insights into a user-friendly and powerful platform for molecular property prediction.

\section{ChemXploreML: A Desktop Application for Molecular Property Prediction}
\label{sec:ChemXploreML}

In this study, we have developed ChemXploreML, a desktop application designed to bridge the gap between ML techniques and everyday chemical research needs. In the following sections, we will briefly describe its design, implementation, and core functionality which are relevant for molecular property predictions using machine learning.

\subsection{Architecture and Technical Implementation}
The application uses a combined software design: one part creates the user interface \cite{tauri2024, svelte2024}, while another handles core computational tasks. The core computational engine is implemented in the Python programming language \cite{python312}, utilizing established scientific computing libraries. This design enables seamless cross-platform compatibility (Windows, macOS, and Linux) and ensures efficient resource utilization. To further assist users and demonstrate the accessibility of the interface, we have developed a public documentation site available at \url{https://aravindhnivas.github.io/ChemXploreML-docs/}. This site includes detailed screenshots, step-by-step usage guides, and installation instructions across platforms.

\subsection{Core Functionality and Features}

ChemXploreML integrates robust data handling with advanced ML implementations in a user-friendly desktop environment. The application's data handling framework supports multiple file formats (CSV, JSON, HDF5) and incorporates extensive molecular analysis through RDKit \cite{greg_landrum_rdkitrdkit_2024} integration. The preprocessing pipeline includes automated scaling, transformation, bootstrapping, and cross-validation procedures essential for robust ML model development.

The machine learning framework combines both traditional and modern approaches. Traditional algorithms include linear models (Linear Regression, Ridge), Support Vector Regression, K-Nearest Neighbors, and tree-based ensembles using \texttt{scikit-learn} \cite{pedregosa_scikit-learn_2011}, while modern boosting frameworks incorporate XGBoost, CatBoost, and LightGBM. Hyperparameter optimization is efficiently handled by Optuna\cite{akiba_optuna_2019}, allowing user-configurable tuning strategies.

Performance is further enhanced by Dask \cite{rocklin_dask_2015} integration, which supports large-scale data processing and configurable parallelization. An intuitive graphical interface provides real-time visualization of data analyses and model performance and supports interactive configuration of models. Users can visualize molecular structures, examine dataset characteristics, and explore structure-property relationships through dynamic plots and statistical analyses. This comprehensive feature set makes sophisticated molecular property prediction accessible while retaining the flexibility required for advanced research. Moreover, the modular architecture facilitates the seamless integration of new features, ensuring adaptability for routine analyses and innovative ML research in chemistry.

In this study, we validate ChemXploreML's capabilities by implementing and evaluating four tree-based ensemble methods: Gradient Boosting Regression (GBR), XGBoost, CatBoost, and LightGBM (LGBM) $-$ to predict five fundamental molecular properties: melting point (MP), boiling point (BP), vapor pressure (VP), critical temperature (CT), and critical pressure (CP). Although the present study focuses on regression tasks, ChemXploreML is model-type agnostic and supports future inclusion of classification workflows. Planned additions include traditional classifiers such as Logistic Regression, Random Forest, and Support Vector Machines, as well as modern methods like XGBoost Classifier, CatBoost Classifier, and shallow neural networks (e.g., multi-layer perceptrons). These additions will expand the platform's utility to a broader range of cheminformatics problems.

\section{Dataset Collection and Description}
\label{sec:data_collection}

\begin{figure*}[!htb]
  \includegraphics[width=1\textwidth,keepaspectratio]{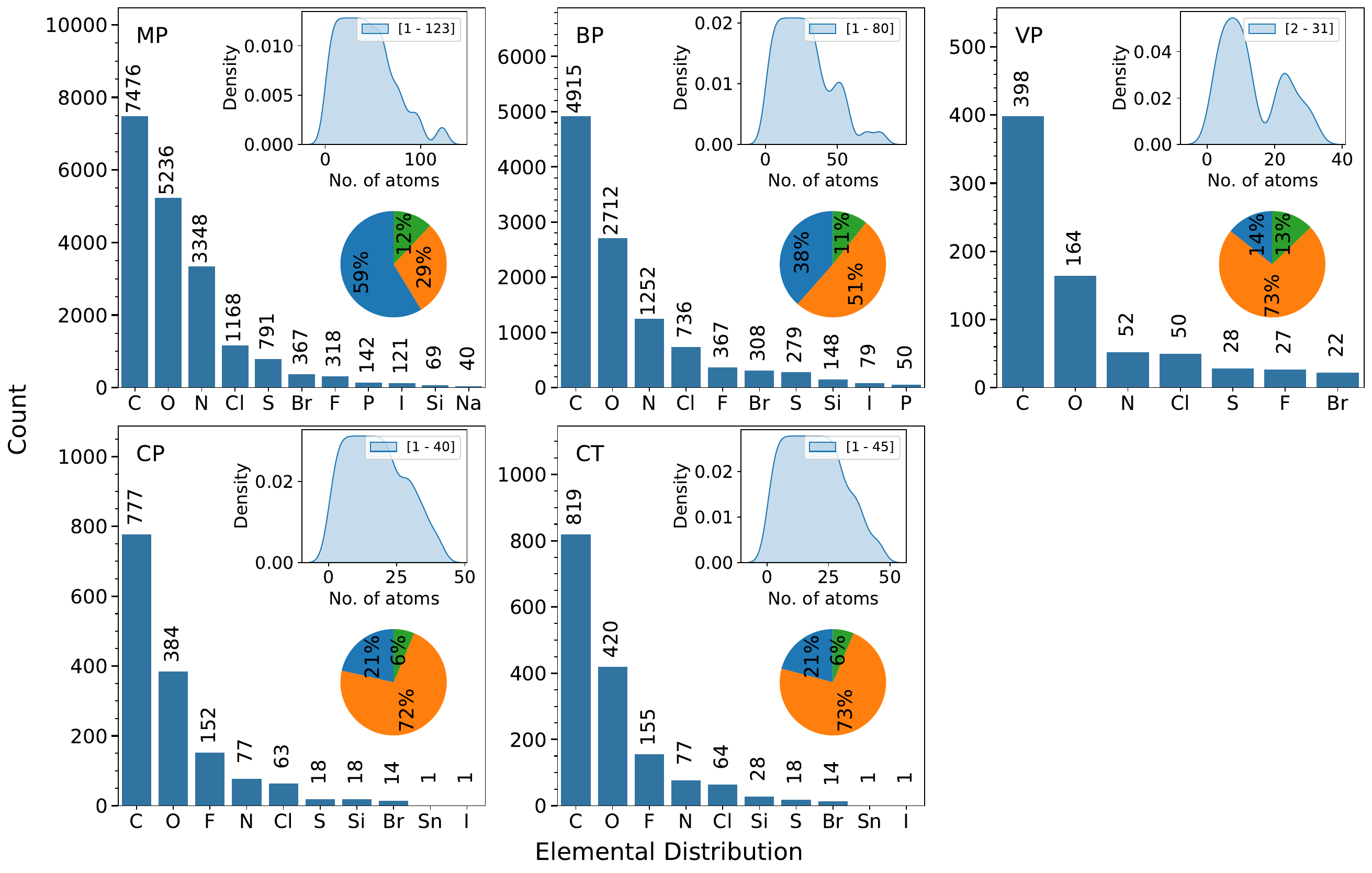}
  \caption{Elemental composition analysis and structural distribution across five thermodynamic property datasets: melting point (MP), boiling point (BP), vapor pressure at $25 ^\circ$C (VP), critical pressure (CP), and critical temperature (CT). Left panels show element frequency distributions, with carbon (C), oxygen (O), and nitrogen (N) being the predominant elements across all datasets. Right panels display molecular size distributions with kernel density estimates, alongside pie charts indicating the structural composition: non-cyclic (orange), aromatic (blue), and cyclic non-aromatic (green) compounds.}
  \label{fig:elemental_distribution}
\end{figure*}

The molecular properties dataset for this study originates from the CRC Handbook of Chemistry and Physics, recognized as a highly reliable and comprehensive reference for chemical and physical properties \cite{rumble_crc_2024}. The dataset includes a diverse range of molecular types, spanning hydrocarbons, halogenated compounds, oxygenated species, and heterocyclic molecules, ensuring broad chemical coverage. The dataset includes five key molecular properties of organic compounds: melting point (MP, $^{\circ}$C), boiling point (BP, $^{\circ}$C), vapor pressure (VP, kPa at 25$^{\circ}$C), critical temperature (CT, K), and critical pressure (CP, MPa). To enable molecular embedding for machine learning tasks, SMILES \cite{weininger_smiles_1988} representations were obtained for each compound using their CAS Registry Numbers. The SMILES strings were primarily retrieved through the PubChem REST API \cite{kim_pug-view_2019}, and when necessary, supplemented using the NCI Chemical Identifier Resolver (CIR) via the cirpy Python interface \cite{noauthor_mcs07cirpy_nodate}

Following SMILES acquisition, we utilized RDKit \cite{greg_landrum_rdkitrdkit_2024} - a leading open-source cheminformatics software package - to canonicalize the SMILES string (i.e., produce a single, standardized representation for each molecule) \cite{oboyle_towards_2012} and extract crucial molecular information. RDKit enabled us to analyze molecular structures systematically, providing detailed insights into atomic composition, connectivity, and structural features as discussed in the following section.

\subsection{Dataset Characteristics and Chemical Space Analysis}

Our study investigates molecular property prediction using comprehensive datasets of organic compounds. Table \ref{tab:dataset_summary} presents the distribution of compounds across different properties, with the melting point dataset being the largest and vapor pressure dataset the smallest. All compounds were represented using canonical SMILES notation, ensuring consistent molecular representation throughout the dataset.

\begin{table}[!htb]
    \centering
    \caption{Summary of molecular properties datasets size. The initial data was sourced from the CRC Handbook of Chemistry and Physics \cite{rumble_crc_2024}, containing experimentally measured properties for melting point (MP, $^{\circ}$C), boiling point (BP, $^{\circ}$C), vapor pressure (VP, kPa at 25$^{\circ}$C), critical temperature (CT, K), and critical pressure (CP, MPa). The Original column shows the initial dataset size, while the Validated column indicates samples with valid molecular embeddings generated by the respective embedders (Mol2Vec and VICGAE). The Cleaned column represents the final dataset size after cleaning data using cleanlab\cite{noauthor_cleanlab_2024, zhou2023errors, kuan2022labelquality}, which identifies and removes potentially problematic data points.}
    \label{tab:dataset_summary}
    \setlength{\tabcolsep}{4pt}
    \begin{tabular}{clcccc}
        \toprule
        Property & Embedder & Original & Validated & Cleaned \\
        \midrule
        \multirow{2}{*}{\text{MP}} & Mol2Vec & 7476 & 7476 & 6167 \\
         & VICGAE & 7476 & 7200 & 6030 \\\\
        \multirow{2}{*}{\text{BP}} & Mol2Vec & 4915 & 4915 & 4816 \\
         & VICGAE & 4915 & 4909 & 4663 \\\\
        \multirow{2}{*}{\text{VP}} & Mol2Vec & 398 & 398 & 353 \\
         & VICGAE & 398 & 398 & 323 \\\\
        \multirow{2}{*}{\text{CP}} & Mol2Vec & 777 & 777 & 753 \\
         & VICGAE & 777 & 776 & 752 \\\\
        \multirow{2}{*}{\text{CT}} & Mol2Vec & 819 & 819 & 819 \\
         & VICGAE & 819 & 818 & 777 \\
        \bottomrule
    \end{tabular}
\end{table}

\begin{figure*}[!htb]
  \includegraphics[width=1\textwidth,keepaspectratio]{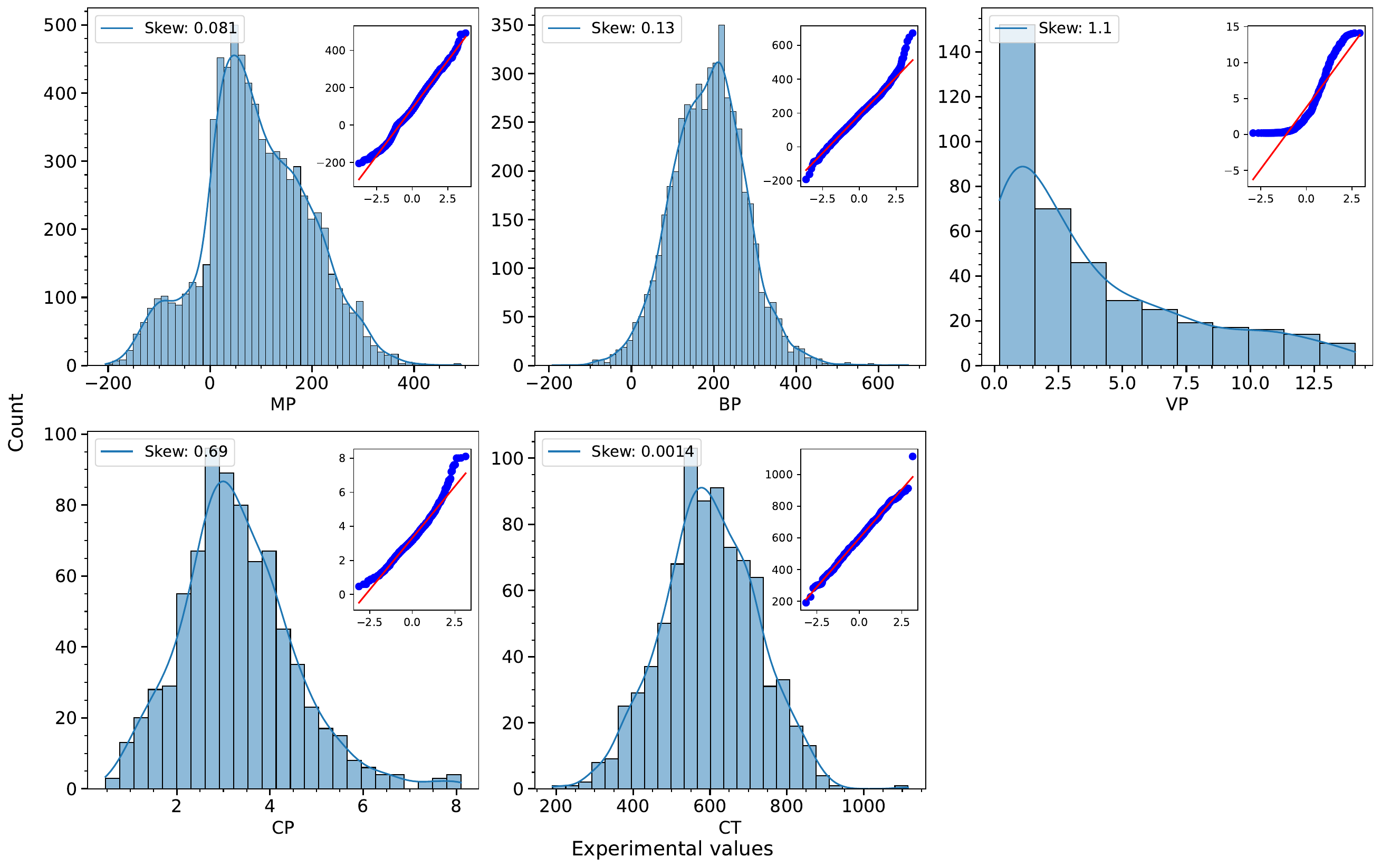}
  \caption{Distribution analysis of molecular properties across the datasets. The main plots show frequency distributions of melting point (MP, $^{\circ}$C), boiling point (BP, $^{\circ}$C), vapor pressure (VP, kPa at 25$^{\circ}$C), critical pressure (CP, MPa), and critical temperature (CT, K). The inset Q-Q plots assess the normality of each distribution, where the linear alignment indicates normal distribution. The calculated skewness values reveal varying degrees of distribution symmetry: MP (0.081), BP (0.13), VP (1.1), CP (0.69), and CT (0.0014), where values closer to zero indicate more symmetric distributions.}
  \label{fig:y_distribution}
\end{figure*}

We used ChemXploreML to conduct an in-depth analysis of our dataset's chemical space and molecular characteristics. ChemXploreML streamlines the chemical space exploration process by providing unified interfaces for elemental distribution analysis, structural classification (aromatic, non-cyclic, and cyclic non-aromatic), and molecular size distribution as shown in Figure \ref{fig:elemental_distribution}. The elemental distribution reveals that the datasets predominantly comprise organic compounds with carbon, oxygen, and nitrogen as the primary constituents. Carbon, being the fundamental building block of organic molecules, is present throughout the dataset. Oxygen emerges as the second most prevalent element, appearing in 5,236 compounds in the MP dataset and 2,712 compounds in the BP dataset. Nitrogen is the third most common element, present in 3,348 and 1,252 compounds in the MP and BP datasets, respectively. Halogens (Cl, F, Br, I) and other elements such as sulfur, phosphorus, and silicon appear in smaller proportions, contributing to the chemical diversity of the dataset.


The molecular complexity analysis, shown in the density plots in (see Figure \ref{fig:elemental_distribution} top inset plot), reveals varying distributions in the data sets. The MP dataset exhibits a balanced distribution with $59\%$ small molecules ($1-40$ atoms), $29\%$ medium-sized molecules ($41-80$ atoms), and $12\%$ large molecules ($>80$ atoms), with molecules ranging from 1 to 123 atoms and a characteristic peak around $25-30$ atoms. The BP dataset shows a slightly different pattern, with $38\%$ small molecules, $51\%$ medium-sized molecules, and $11\%$ large molecules, exhibiting a narrower range ($1-80$ atoms) with a similar peak distribution. The VP dataset predominantly contains medium-sized molecules ($73\%$) with the most constrained size distribution ($2-31$ atoms), while both CT and CP datasets show similar distributions with approximately $72-73\%$ medium-sized molecules, ranging from 1 to approximately $40-45$ atoms.

The structural composition of the datasets, illustrated in the pie charts in Figure \ref{fig:elemental_distribution}, provides crucial insights into molecular structural distribution. For the MP dataset, non-cyclic compounds constitute 59\% of the molecules, while aromatic and non-aromatic cyclic structures account for 29\% and 12\%, respectively. The BP dataset shows a different distribution with 38\% non-cyclic compounds, a higher proportion of aromatic structures at 51\%, and 11\% non-aromatic cyclic compounds. Notably, the VP dataset displays a distinct pattern with 73\% aromatic compounds. The CT and CP datasets exhibit similar structural distributions, with approximately $72-73\%$ aromatic compounds.

\subsection{Property Distribution Analysis}
\label{sec:property_distribution_analysis}

Understanding the distribution of molecular properties is crucial for selecting appropriate ML models and interpreting their predictions. Figure \ref{fig:y_distribution} visualizes the distribution of each property using histograms and corresponding Q-Q plots as insets. Q-Q plots (quantile-quantile plots) compare the quantiles of the data distribution to the quantiles of a theoretical normal distribution.  A linear Q-Q plot indicates that the data follows a normal distribution, while deviations from linearity suggest departures from normality.

Melting point and boiling point exhibit similar distributional characteristics. Melting point values follow an approximately normal distribution with minimal skewness (0.081), as evidenced by the linear Q-Q plot. This indicates that the data is evenly distributed around the mean. Similarly, the boiling point shows a slight positive skew (0.13), suggesting a slightly longer tail on the right side of the distribution, indicating a few compounds with unusually high boiling points. Vapor pressure exhibits the most pronounced positive skew (1.1), indicating the presence of compounds with exceptionally high vapor pressures compared to the majority of the dataset. 

Critical pressure demonstrates moderate positive skewness (0.69), and critical temperature shows an almost perfectly symmetric distribution (skewness: 0.0014), with its Q-Q plot showing excellent alignment with the theoretical normal distribution.


\section{Molecular Representation}
\label{sec:molecular_representation}

Machine learning (ML) algorithms require numerical representations of molecular structures. While traditional descriptors and fingerprints have been widely used \cite{riniker_open-source_2013, oboyle_comparing_2016, riniker_heterogeneous_2013, mayr_deeptox_2016, merget_profiling_2017, sorgenfrei_kinome-wide_2018}, recent embedding techniques capture more nuanced chemical information by mapping molecules into continuous vector spaces. In this work, we implemented two state-of-the-art embeddings, Mol2Vec and VICGAE, which were chosen for their ability to encode local (functional group-level) and global (structural/conformational) features.

\subsection{Mol2Vec Embedding Method}
Mol2Vec \cite{jaeger_mol2vec_2018} is an unsupervised molecular embedding method inspired by natural language processing (specifically Word2Vec \cite{mikolov_efficient_2013}). In Mol2Vec, molecules are \say{parsed} into fragments (e.g. Morgan algorithm atom-centered substructures \cite{rogers_extended-connectivity_2010}), which play the role of words. A skip-gram neural network \cite{mikolov_efficient_2013} is then trained on a large corpus of molecules to learn vector representations for these substructure “words,” such that fragments appearing in similar chemical contexts have similar embeddings. Each molecule’s vector is obtained by summing its fragment embeddings, yielding a fixed-length feature (typically 300 dimensions as used by \citet{jaeger_mol2vec_2018}).
These pre-trained Mol2Vec vectors, accessible via SMILES, has shown strong performance in drug repurposing \cite{das_repurposed_2021} and property prediction \cite{zheng_identifying_2019, lee_machine_2021, fried_implementation_2023}. The advantage of Mol2Vec is that it requires no explicit labeling of data for training and can leverage large unlabeled databases to create informative features. Because it encodes local chemical environments, it captures functional group-level information, which can be complementary to global structure features. Mol2Vec’s effectiveness has been validated in a variety of regression tasks, often improving accuracy when used in place of or alongside traditional fingerprints \cite{lee_machine_2021, fried_implementation_2023}.

\subsection{VICGAE Embedding Method}
In a recent advancement in molecular representation learning, VICGAE (Variance-Invariance-Covariance regularized GRU Auto-Encoder) offers a sophisticated approach to generating compact molecular embeddings. Introduced by \citet{scolati_explaining_2023} in the context of astrochemistry to embed molecules for which ensuring similar embeddings for chemically related species is important. This deep generative model processes SELFIES (\textbf{SELF}-referenc\textbf{I}ng \textbf{E}mbedded \textbf{S}trings) \cite{krenn_self-referencing_2020} string representations through a sequence-to-sequence autoencoder, training on partially masked sequences to reconstruct missing tokens. By incorporating VIC regularization \cite{bardes_vicreg_2022}, VICGAE ensures embeddings exhibit high \textit{variance} for effective dimensional use, \textit{invariance} to trivial transformations (e.g., isotopic substitutions), and low \textit{covariance} between dimensions, preventing latent space collapse. The resulting 32-dimensional vectors adeptly capture both global structural features and subtle chemical variations, making VICGAE ideal for representing chemically related species, such as isotopologues, with similar embeddings, thus enhancing chemical similarity analysis.

\subsection{Other Embedding Methods}

\textbf{Graph Neural Networks (GNNs)}, such as the AttentiveFP GCN, excel at predicting molecular properties like toxicity and bioactivity, especially with large datasets \cite{xiong_pushing_2020}. They outperform older graph models and fingerprint-based neural networks, delivering notable improvements in accuracy metrics like ROC-AUC and RMSE. However, their edge over simpler methods fades when data is scarce, as they thrive best with complex structure-property relationships and ample training data \cite{yang_analyzing_2019, yang_correction_2019}.

\textbf{Transformer Models}, including SMILES2Vec RNNs \cite{goh_smiles2vec_2018} and transformers like ChemBERTa\cite{chithrananda_chemberta_2020} and MolBERT\cite{li_mol-bert_2021}, are strong contenders in molecular property prediction. The standout MoLFormer-XL, pre-trained on 1.1 billion molecules, achieves top accuracy on numerous benchmarks, even surpassing GNNs with 3D geometry \cite{ross_large-scale_2022}. Yet, simpler approaches (GBFS-Mol2Vec \cite{jung_gradient_2023}) can sometimes match or outperform these complex models, particularly when the latter are trained on smaller datasets, showing that performance varies by context \cite{jung_automatic_2024}. This reinforces that no one method is universally best, and performance depends on the context (data size, property type, etc.).

\subsection{Implementation in ChemXploreML}

As discussed above, recent literature suggests that machine-learned molecular embeddings (Mol2Vec, graph-based, SMILES-based, etc.) generally improve prediction performance over fixed fingerprints, especially as task complexity grows. Each approach has its strengths: Mol2Vec excels in simplicity and leveraging big data unsupervised, VICGAE provides compact yet information-rich vectors, GCNs capture detailed structural relationships, and SMILES transformers benefit from massive data training. In head-to-head comparisons on property prediction, differences in accuracy can be property-dependent. 


Therefore, ChemXploreML currently supports Mol2Vec (300-D) and VICGAE (32-D) embedders, which are evaluated in this study to provide a robust and interpretable baseline for structure–property exploration (see Section~\ref{sec:regression_models}). Beyond these, ChemXploreML is built to be modular and easily extensible. We have recently integrated two additional embeddings—ChemBERTa and MoLFormer—both of which are now available within the application. These transformer-based models offer additional flexibility for users seeking alternative representation learning strategies. Although the results from these newer embeddings are not analyzed in the present work, they will be considered in future benchmarking efforts. The system architecture further allows integration of graph neural networks (e.g., GCN, AttentiveFP) and 3D-aware models as needed for advanced applications.

ChemXploreML also includes a diverse set of dimensionality reduction tools, including PCA, KernelPCA, t-SNE, PHATE, ISOMAP, Laplacian Eigenmaps, TriMap, and Factor Analysis. While UMAP was exclusively used in this study due to its strength in capturing nonlinear structure–property trends, the interface allows users to select from multiple DR methods for exploratory analysis. Future work could leverage these tools to benchmark dimensionality reduction techniques in cheminformatics settings.


\section{Mapping Chemical Space with UMAP}
\label{sec:umap}

\begin{figure*}[!htb]
  \includegraphics[width=1\textwidth,keepaspectratio]{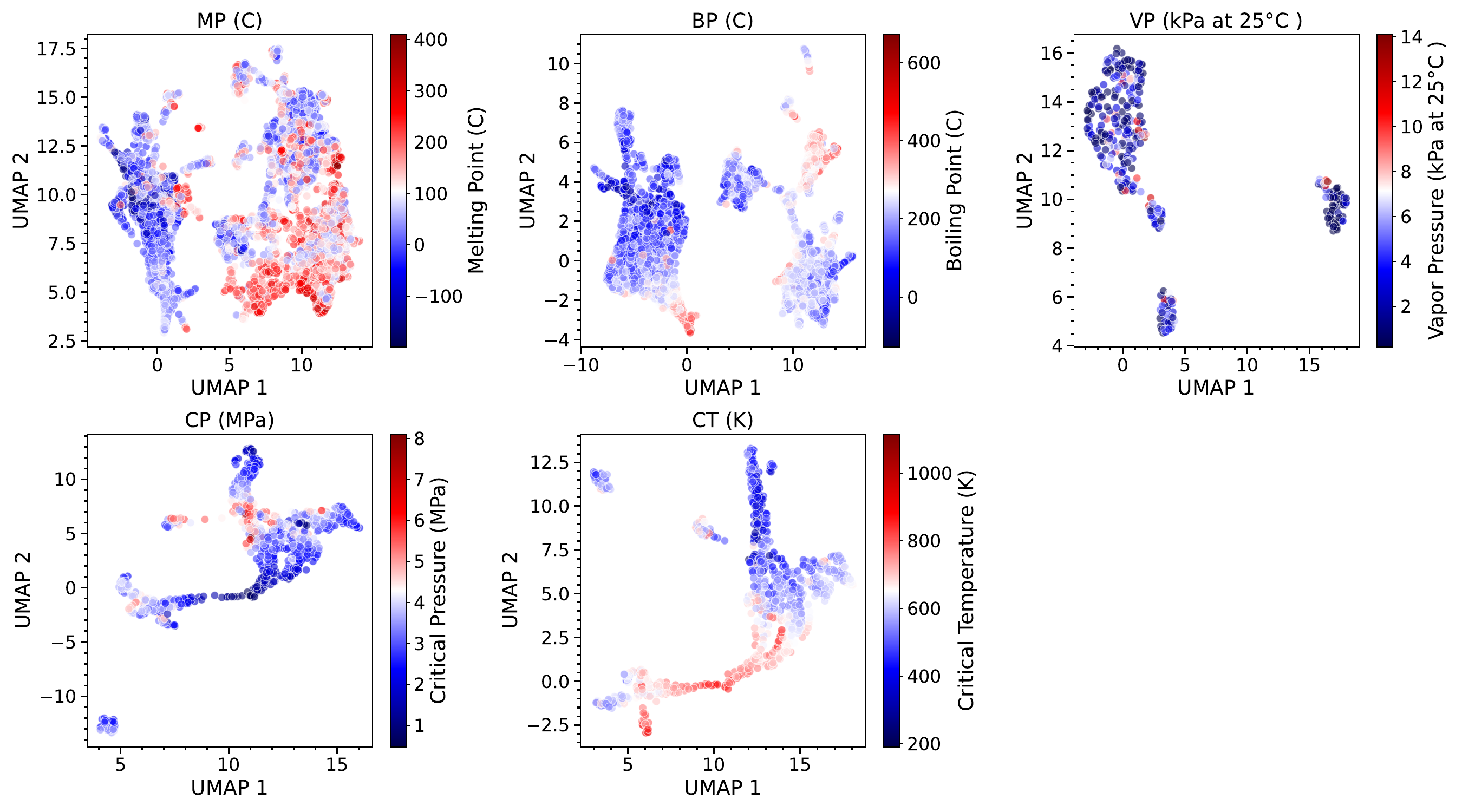}
  \caption{UMAP projections illustrating the clustering patterns of molecular embeddings for five distinct properties: melting point (MP, $^{\circ}$C), boiling point (BP, $^{\circ}$C), vapor pressure (VP, kPa at 25$^{\circ}$C), critical pressure (CP, MPa), and critical temperature (CT, K). Each subplot presents a two-dimensional representation of the high-dimensional molecular data, with points colored according to their respective property values. Note: The colormap represented here was chosen due to its high visual contrast.  A version of this figure with a perceptually uniform colormap is additionally provided in the SI (Figure S1).
  }
  \label{fig:umap}
\end{figure*}

Understanding the relationship between molecular structures and their properties is a core challenge in chemistry. Uniform Manifold Approximation and Projection (UMAP) facilitates the visualization of high-dimensional molecular embeddings—such as those from Mol2Vec—by mapping them into a lower-dimensional, i.e., 2D space, while preserving structural relationships. 

While Principal Component Analysis (PCA) is a commonly used linear technique, it may not adequately capture the nonlinear structure–property relationships present in molecular embedding spaces. In contrast, UMAP preserves both local and global topological features, making it more suitable for revealing meaningful clustering patterns in complex, high-dimensional chemical data. Although UMAP was the primary method employed for unsupervised mapping in this study, future work may explore comparative assessments of different dimensionality reduction techniques for cheminformatics tasks.

UMAP has two key parameters that significantly influence the quality of the projection: \textit{n\_neighbors}  which controls the balance between local and global structure retention, and \textit{min\_dist}, which determines how closely points are packed in the reduced space. A direct optimization of these parameters is nontrivial, as metrics like trustworthiness do not effectively guide selection. Instead, we optimized UMAP parameters empirically (\textit{n\_neighbors}=25, \textit{min\_dist}=0.3) through visual assessments of molecular structure, ring composition (using RDKit \cite{greg_landrum_rdkitrdkit_2024}), and functional group distributions (using checkmol \cite{haider_functionality_2010}). DBSCAN (Density-Based Spatial Clustering of Applications with Noise) \cite{ester_density-based_1996, schubert_dbscan_2017} clustering (\textit{eps}=0.7, \textit{min\_samples}=15) was similarly refined through structural and chemical analyses.

In Figure \ref{fig:umap}, we present 2D UMAP projections of the Mol2Vec embeddings for our dataset, with each point representing a molecule color-coded by its respective molecular property. This visualization reveals distinct clusters, indicating that molecules with similar structural features and properties are grouped together. Here, we have deliberately used the colormap (\texttt{seismic}), which is not perceptually uniform, to highlight the contrast in property values. For comparison, a perceptually uniform version of this plot is provided in the Supplementary Information (Figure S1).

It is important to note that the UMAP axes do not correspond to explicit chemical properties but instead reflect non-linear combinations of structural similarities embedded in high-dimensional space. To extract chemical insights from this abstract layout, we conducted a detailed post hoc analysis using DBSCAN clustering and functional group annotation (see Supporting Information, Figures S4–S18).

As shown in the figures in supplementary Section S3 (refer to Supporting Information Figures S4–S18), the thermodynamic properties of organic compounds, including melting point (MP) and boiling point (BP), exhibit strong dependencies on molecular structure, intermolecular interactions, and functional group composition. Using unsupervised molecular embedding via UMAP and DBSCAN clustering, we identified distinct molecular distributions that correlate with these properties. MP trends align with molecular weight, hydrogen bonding capacity, and functional group composition, as evident from clustering patterns where high-MP compounds are enriched in carboxylates and polyfunctionalized oxygenated species, whereas low-MP clusters predominantly contain alkanes and simple heterocycles. BP distributions follow expected trends with molecular weight and polarity, with deviations observed for branched structures and conjugated systems that influence dispersion forces and dipole interactions. VP, inversely related to BP, is significantly higher in fluorinated and ether-containing molecules due to their weak intermolecular interactions, while strong hydrogen bonding and extended conjugation reduce VP by increasing molecular cohesion, as illustrated by VP distribution plots across clusters.

Critical temperature (CT) and critical pressure (CP) exhibit structured clustering patterns that align with molecular size and intermolecular interaction strength. CT clustering reveals a clear stratification where polycyclic and conjugated systems demonstrate higher CT values due to delocalized electron density stabilizing the supercritical phase, whereas sterically hindered alkanes and branched hydrocarbons cluster at lower CT due to reduced van der Waals interactions. CP trends highlight the role of molecular compactness and intermolecular repulsions, with small hydrocarbons exhibiting higher CP due to their limited intermolecular interaction, while oxygenated functionalities systematically lower CP by enhancing dipolar interactions. These findings, supported by UMAP clustering and functional group distribution analyses, underscore the utility of unsupervised learning in characterizing thermodynamic trends in organic compounds, offering insights into molecular design for applications in fundamental chemistry and materials science.

These observations underscore the remarkable capacity of unsupervised learning methods to capture complex structure–property relationships that align with established chemical principles. The patterns revealed by UMAP and DBSCAN echo our understanding of how structural stability, ring-structure composition, hydrogen bonding and functional group composition influence key molecular properties. However, as noted earlier, realizing this potential required meticulous empirical optimization of UMAP parameters guided by visual and chemical assessments rather than conventional metrics like trustworthiness.

The next section describes how our preprocessed datasets were refined and prepared for the regression tasks that follow. 

\section{Data Preprocessing Pipeline}
\label{sec:data_preprocessing_pipeline}

Our preprocessing pipeline consists of several key stages designed to handle chemical data robustly while maintaining both high quality and reproducibility. As discussed in Section \ref{sec:data_collection}, initial data validation focuses on ensuring structural integrity through comprehensive analysis using RDKit \cite{greg_landrum_rdkitrdkit_2024}. This includes the removal of \text{invalid} SMILES strings, SMILES canonicalization and elimination of duplicates (using CanonicalSMILES). A thorough molecular characterization analysis enables systematic filtering based on molecular size and elemental composition, as illustrated in Figure \ref{fig:elemental_distribution} which shows the elemental distribution of the analyzed compounds.

The validated structures then undergo molecular embedding generation using two complementary approaches described in Section \ref{sec:molecular_representation}. Notably, Mol2Vec demonstrates exceptional robustness, successfully generating valid 300-dimensional embeddings for 100\% of the input structures across all datasets. In contrast, VICGAE shows slightly lower but still impressive validation rates, successfully embedding $96-99\%$ of the input structures into 32-dimensional vectors across different property datasets (see Table \ref{tab:dataset_summary}).

Post-embedding, for automated outlier detection, we leverage cleanlab \cite{noauthor_cleanlab_2024, zhou2023errors, kuan2022labelquality} to identify and remove problematic data points. Compared to traditional statistical outlier detection (e.g., Z-score filtering, Mahalanobis distance), cleanlab’s confident learning framework better captures mislabeled or inconsistent samples while preserving chemically meaningful variations. The cleanlab framework employs confident learning algorithms to detect out-of-distribution samples and estimate label noise probabilities. The ensemble-based pruning approach ensures robust outlier removal while preserving dataset diversity.
As shown in Table \ref{tab:dataset_summary}, while the melting point dataset experienced an 18\% reduction in Mol2Vec embeddings, other properties—such as BP, CP and VP —retained 98\%, 97\% and 89\% of validated samples, respectively, with no potential outlier detected for the CT dataset (100\% retained), demonstrating that only a minimal fraction of data is pruned, which inherently enhances data reliability for robust model training.

All preprocessing steps are automated within ChemXploreML's workflow, ensuring reproducible data preparation. The preprocessed datasets, containing only validated embeddings and their corresponding property values, form the foundation for the regression models detailed in Section \ref{sec:regression_models}. This rigorous preprocessing approach balances the need for data quality with the preservation of chemical diversity essential for robust model development. Having established a robust preprocessing strategy, we now turn our attention to the regression models used for property prediction in the next section.

\section{Regression Models}
\label{sec:regression_models}

For predicting molecular properties, we employed four state-of-the-art tree-based ensemble methods: Gradient Boosting Regression (GBR), XGBoost, CatBoost, and LightGBM. These models were selected because of their proven effectiveness at handling complex molecular data and their strong capability for capturing non-linear relationships between molecular features and target properties \cite{boldini_practical_2023}.

\subsection{Hyperparameter Optimization}
\label{sec:optuna}

We performed hyperparameter optimization for all models using Optuna, a framework that employs efficient search algorithms to identify the best model configurations \cite{akiba_optuna_2019}. For each model, we defined specific parameter space grid based on their characteristics as given in the Table \ref{tab:hyperparams_grid}.

\begin{table}[!ht]
    \footnotesize
    \caption{Hyperparameter optimization grid space}
    \label{tab:hyperparams_grid}
    \begin{tabular}{lllll}
        \hline
        Model & Parameter & Range \\
        \hline
        \hline
        \multirow{8}{*}{GBR} \\
            & \emph{n\_estimators} & $100-1000$ \\
            & \emph{learning\_rate} & $10^{-3}-1.0$ \\
            & \emph{max\_depth} & $1-10$ \\
            & \emph{min\_samples\_split} & $2-20$ \\
            & \emph{min\_samples\_leaf} & $1-10$ \\
            & \emph{subsample} & $0.5-1.0$ \\
        \hline
        \multirow{8}{*}{XGBoost} \\
            & \emph{n\_estimators} & $100-1000$ \\
            & \emph{learning\_rate} & $10^{-3}-1.0$ \\
            & \emph{max\_depth} & $1-10$ \\
            & \emph{colsample\_bytree} & $0.5-1.0$ \\
            & \emph{min\_child\_weight} & $1-10$ \\
            & \emph{subsample} & $0.5-1.0$ \\
        \hline
        \multirow{8}{*}{CatBoost} \\
            & \emph{iterations} & $100-1000$ \\
            & \emph{learning\_rate} & $10^{-3}-1.0$ \\
            & \emph{depth} & $4-10$ \\
            & \emph{l2\_leaf\_reg} & $10^{-8}-100.0$ \\
            & \emph{random\_strength} & $10^{-8}-100.0$ \\
            & \emph{subsample} & $0.5-1.0$ \\
        \hline
        \multirow{8}{*}{LightGBM} \\
            & \emph{n\_estimators} & $100-1000$ \\
            & \emph{num\_leaves} & $20-3000$ \\
            & \emph{learning\_rate} & $10^{-3}-1.0$ \\
            & \emph{colsample\_bytree} & $0.5-1.0$ \\
            & \emph{min\_child\_samples} & $5-100$ \\
            & \emph{subsample} & $0.5-1.0$ \\
        \hline
    \end{tabular}
\end{table}

GBR's optimization space included the number of estimators ($100-1000$), learning rate ($10^{-3}-1.0$), maximum depth ($1-10$), and various sampling parameters to control the model's complexity and prevent overfitting. XGBoost was tuned with similar base parameters along with its unique features such as column sampling by tree ($0.5-1.0$) and minimum child weight ($1-10$).

CatBoost's parameter grid encompassed iterations ($100-1000$), learning rate ($10^{-3}-1.0$), depth ($4-10$), and regularization parameters including L2 leaf regularization and random strength (both ranging from $10^{-8}$ to $100.0$). LightGBM's optimization focused on the number of estimators ($100-1000$), number of leaves ($20-3000$), learning rate ($10^{-3}-1.0$), and sampling parameters for both columns and data points. All models incorporated subsampling strategies ($0.5-1.0$) to enhance generalization and prevent overfitting. 

\subsection{N-fold Cross validation}
\label{sec:cross_validation}

The optimization process employed a $5-$fold (N$=5$) cross-validation (CV) strategy to ensure robust performance estimates across different data splits. In N-fold CV, the dataset is divided into N equal-sized partitions, where N-1 partitions are used for training and the remaining partition serves as the validation set. This process is repeated N times, with each partition serving as the validation set exactly once, and the performance metrics are averaged across all N iterations. This approach provides a more reliable estimate of the model performance by ensuring that each data point appears in both the training and the validation sets. This comprehensive hyperparameter tuning approach ensured that each model achieved its optimal configuration for the specific molecular property prediction tasks.

With these regression models fine-tuned, we now discuss the results of our predictive experiments and key insights gained from their performance.

\section{Results and Discussion}
\label{sec:results_and_discussion}

We performed a comprehensive evaluation of molecular property prediction using four machine learning algorithms combined with two molecular embedding approaches (as discussed in Section \ref{sec:data_preprocessing_pipeline} and \ref{sec:regression_models}). The models were assessed on five molecular properties: melting point (MP), boiling point (BP), vapor pressure (VP), critical temperature (CT), and critical pressure (CP). In this section, we analyze the results and performance of the ML models and embeddings for the aforementioned molecular properties.

\subsection{Hyperparameter Optimization Analysis}

We employed Optuna \cite{akiba_optuna_2019}, a hyperparameter optimization framework that utilizes Tree-structured Parzen Estimators (TPE) \cite{watanabe_tree-structured_2023}, for efficient parameter tuning. Unlike traditional methods like GridSearchCV or RandomSearchCV, Optuna implements a Bayesian optimization \cite{snoek_practical_2012} approach that adaptively samples the parameter space based on previous trial results, leading to faster convergence and better parameter combinations. Optuna's ability to handle conditional parameters and prune unpromising trials makes it particularly suitable for complex molecular property prediction models where parameter interactions significantly impact performance.

\begin{figure}[!htb]
  \includegraphics[width=0.5\textwidth,keepaspectratio]{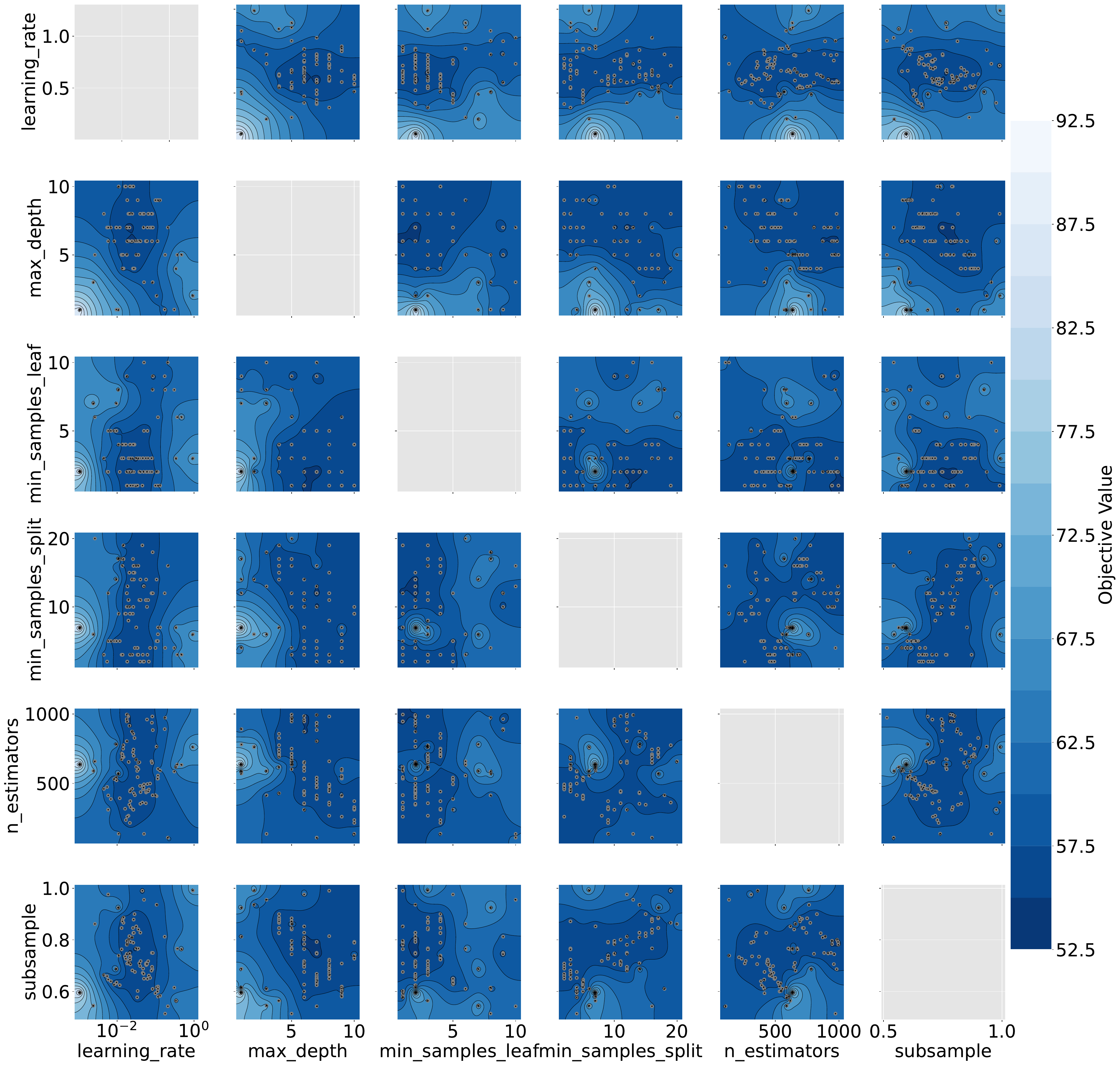}
  \caption{Hyperparameter optimization landscape for MP with Gradient Boosting Regressor using Mol2Vec embeddings, visualized through Optuna's contour plot. The optimization space explores key GBR parameters. Darker regions represent more optimal parameter combinations that minimize the objective function (minimizing RMSE). The gray dots in each subplot represent the actual hyperparameter combinations that were sampled}
  \label{fig:MP_Optuna_contour}
\end{figure}

\begin{table*}[!htb]
    \centering
    \caption{Optimized hyperparameters for different regression models across molecular properties}
    \scriptsize
    \begin{tabular}{llccccc}
        \toprule
        Model & Parameter & MP & BP & VP & CP & CT \\
        \midrule
        \multirow{6}{*}{GBR} 
        & \emph{n\_estimators} & 804, 633 & 939, 986 & 921, 1000 & 513, 958 & 929, 509 \\
        & \emph{learning\_rate} & 0.02, 0.02 & 0.05, 0.02 & 0.01, 0.07 & 0.04, 0.05 & 0.01, 0.03 \\
        & \emph{min\_samples\_split} & 12, 4 & 6, 20 & 15, 6 & 12, 14 & 17, 20 \\
        & \emph{min\_samples\_leaf} & 2, 5 & 9, 2 & 7, 5 & 6, 4 & 9, 7 \\
        & \emph{max\_depth} & 7, 8 & 5, 7 & 3, 3 & 4, 4 & 4, 9 \\
        & \emph{subsample} & 0.7, 0.6 & 0.5, 0.7 & 0.6, 1.0 & 0.9, 0.5 & 0.5, 0.5 \\
        \midrule
        \multirow{6}{*}{XGBOOST} & n\_estimators & 550, 644 & 822, 918 & 627, 875 & 887, 462 & 676, 144 \\
        & \emph{learning\_rate} & 0.3, 0.3 & 0.3, 0.3 & 0.3, 0.3 & 0.3, 0.3 & 0.5, 0.5 \\
        & \emph{max\_depth} & 7, 8 & 10, 9 & 9, 8 & 9, 8 & 8, 10 \\
        & \emph{subsample} & 1.0, 1.0 & 0.9, 1.0 & 0.9, 0.6 & 0.8, 0.7 & 0.9, 0.7 \\
        & \emph{colsample\_bytree} & 0.8, 0.9 & 0.9, 0.5 & 0.6, 0.7 & 0.6, 0.7 & 1.0, 0.9 \\
        & \emph{min\_child\_weight} & 2, 4 & 4, 5 & 8, 5 & 3, 5 & 7, 8 \\
        \midrule
        \multirow{5}{*}{CATBOOST} & iterations & 627, 497 & 426, 722 & 621, 792 & 609, 711 & 856, 922 \\
        & \emph{learning\_rate} & 0.1, 0.3 & 0.2, 0.1 & 0.2, 0.2 & 0.3, 0.2 & 0.2, 0.2 \\
        & \emph{depth} & 3, 9 & 11, 8 & 2, 13 & 3, 5 & 9, 8 \\
        & \emph{random\_strength} & 0.1, 0.3 & 1.69E-07, 1.40E-05 & 0.03, 1.22E-07 & 0.29, 0.02 & 9.90E-07, 2.39E-04 \\
        & \emph{subsample} & 0.8, 0.7 & 0.8, 0.7 & 0.7, 1.0 & 1.0, 0.6 & 0.8, 0.8 \\
        \midrule
        \multirow{6}{*}{LGBM} & n\_estimators & 579, 853 & 480, 530 & 517, 792 & 581, 149 & 239, 948 \\
        & \emph{learning\_rate} & 0.03, 0.02 & 0.1, 0.1 & 0.2, 0.4 & 0.2, 0.4 & 0.2, 0.2 \\
        & \emph{subsample} & 0.7, 0.8 & 0.9, 0.9 & 0.7, 0.8 & 0.7, 0.8 & 0.7, 0.8 \\
        & \emph{colsample\_bytree} & 1.0, 0.6 & 0.8, 0.6 & 0.6, 0.7 & 0.5, 0.9 & 0.6, 0.7 \\
        & \emph{num\_leaves} & 251, 101 & 120, 167 & 156, 230 & 114, 4 & 163, 46 \\
        & \emph{min\_child\_samples} & 97, 5 & 60, 21 & 14, 35 & 57, 14 & 15, 60 \\
        \midrule
        \bottomrule\\
    \end{tabular}
    \label{tab:hyperparams_optimized}

    \small{Note: MP = Melting Point, BP = Boiling Point, VP = Vapor Pressure, CP = Critical Pressure, CT = Critical Temperature. For each property, two values are shown representing Mol2vec and VICGAE embedder results respectively.}
\end{table*}

For each model, we performed extensive hyperparameter tuning using Optuna with 5-fold cross-validation to ensure robust performance. The optimization landscape for the regression models reveals interesting patterns across different molecular properties. The GBR model's optimization landscape for the MP analysis is effectively visualized in Figure \ref{fig:MP_Optuna_contour} as a contour plot. The contour plot matrix shows the relationships between six key hyperparameters optimized for GBR: \emph{learning\_rate, max\_depth, min\_samples\_leaf, min\_samples\_split, n\_estimators,} and \emph{subsample}. The gray dots represent sampled parameter combinations, while contour colors indicate performance (darker regions showing better performance i.e., lower RMSE). Table \ref{tab:hyperparams_optimized} provides the optimized values for these parameters across different molecular properties.

The density and distribution of these gray dots in Figure \ref{fig:MP_Optuna_contour} (contour plots) provide important insights.

\begin{enumerate}
    \item Areas with concentrated dots indicate regions where the optimization algorithm focused its search, suggesting promising performance areas.
    \item The varying density of dots across the parameter space shows how Optuna adaptively sampled different regions based on previous trial results.
    \item Sparser dots in certain regions typically indicate areas where the algorithm determined lower performance potential and reduced exploration.
\end{enumerate}

This sampling pattern visualization is particularly valuable for understanding how the optimization algorithm explored the parameter space and ultimately converged on the optimal configurations reported in Table \ref{tab:hyperparams_optimized}. For example, in the \emph{learning\_rate} vs. \emph{n\_estimators} plot, we can see a higher concentration of samples in the darker regions where performance was better, demonstrating the algorithm's efficient focus on promising parameter combinations.

The comprehensive hyperparameter optimization results demonstrate the importance of careful model tuning for molecular property prediction tasks. These optimized configurations serve as the foundation for comparing model performances across different molecular properties and embedding techniques.

\subsection{Performance Analysis of Machine Learning Models and Embeddings}

\begin{figure}[!htb]
  \includegraphics[width=\columnwidth,keepaspectratio]{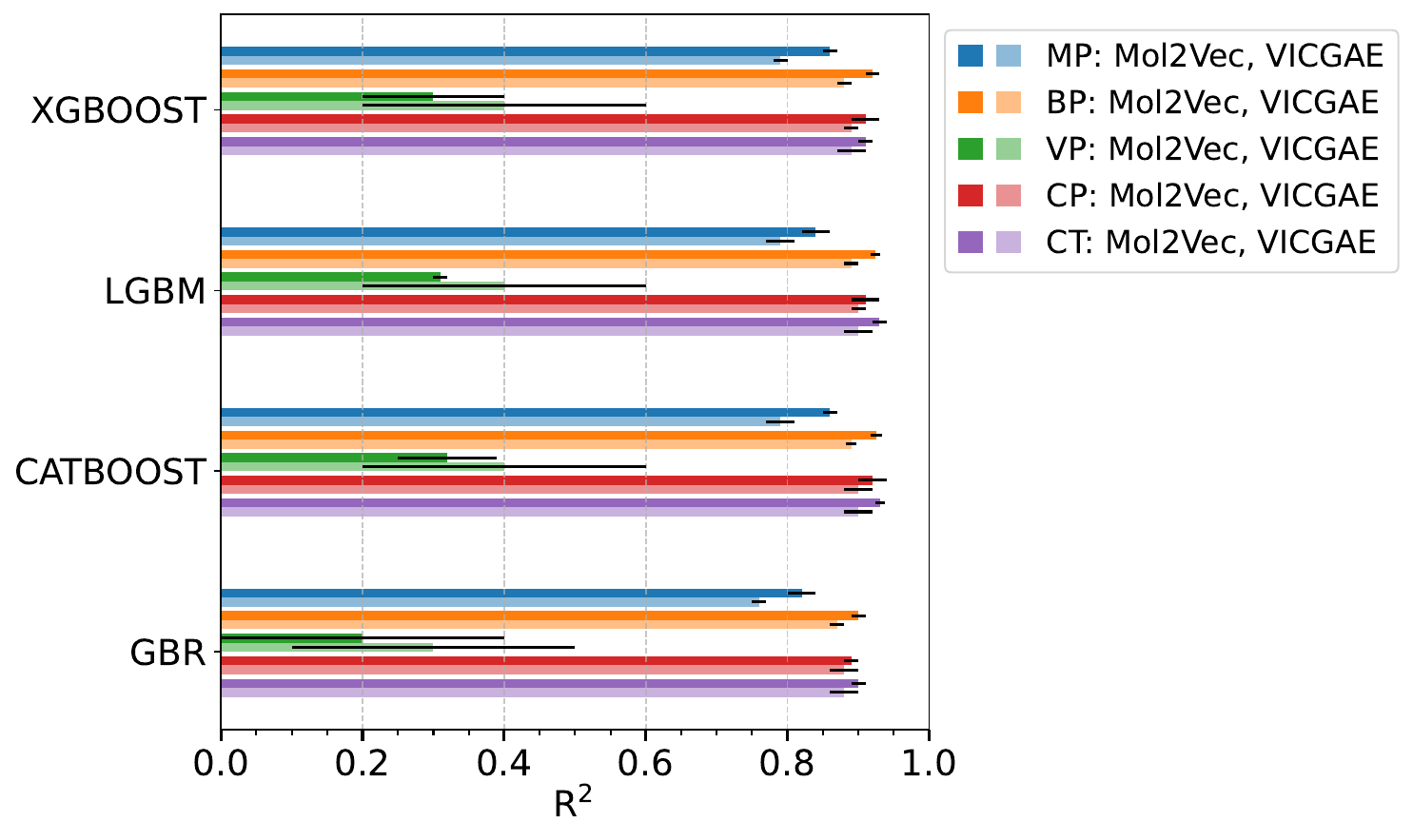}
  \caption{Performance comparison of machine learning models across different molecular properties using both Mol2Vec and VICGAE embeddings. The R$^2$ scores are shown for GBR, CatBoost, LGBM, and XGBoost models predicting five molecular properties: melting point (MP, $^{\circ}$C), boiling point (BP, $^{\circ}$C), vapor pressure (VP, kPa at 25$^{\circ}$C), critical temperature (CT, K), and critical pressure (CP, MPa).  For each property, solid bars represent predictions using Mol2Vec embeddings while lighter bars with the same color indicate VICGAE embeddings. The black solid lines represent error bars (standard deviation) obtained through 5-fold CV.}
  \label{fig:r2_scores}
\end{figure}

\begin{figure*}[!htb]
  \includegraphics[width=\textwidth,keepaspectratio]{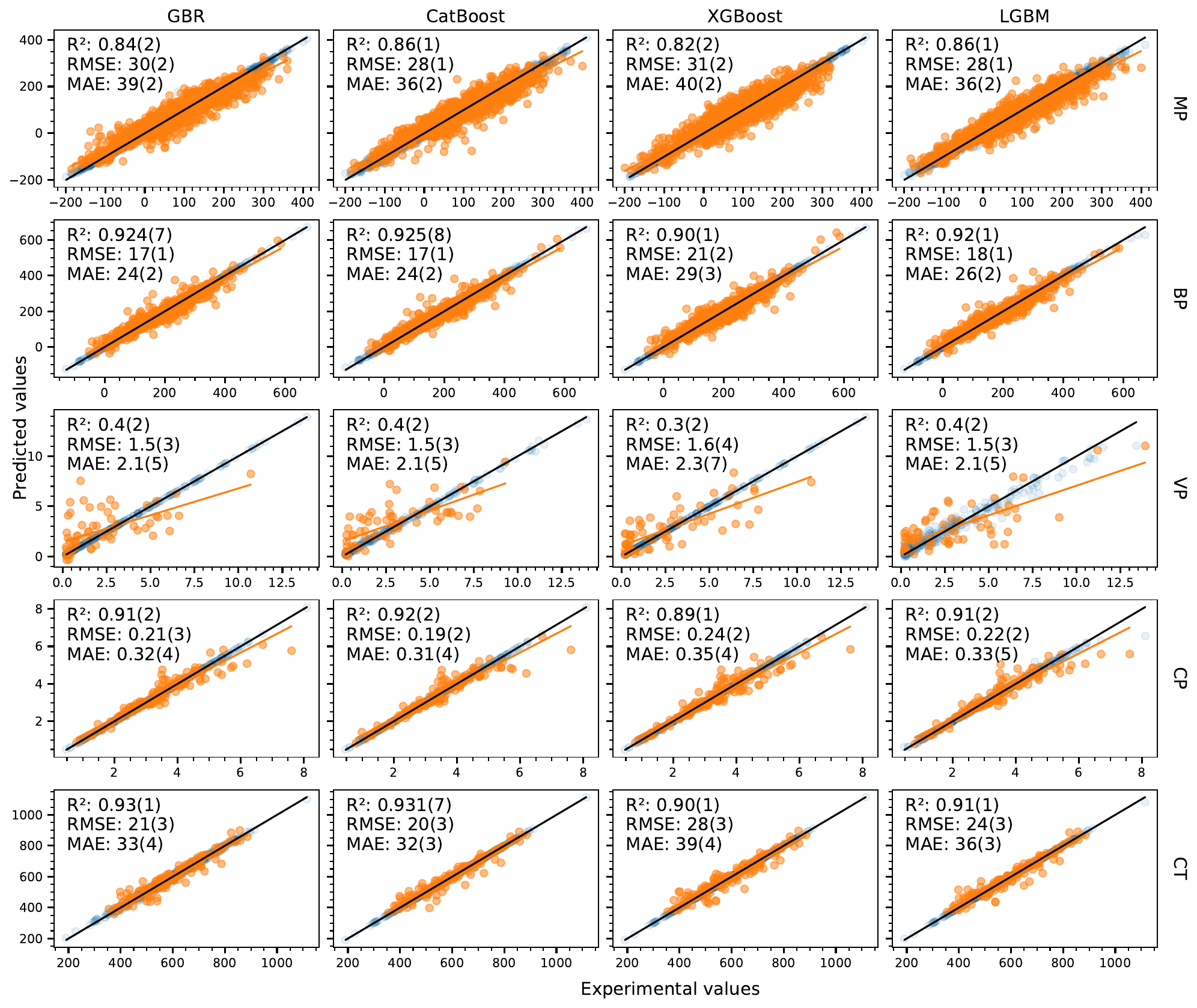}
  \caption{Comparison of experimental versus predicted values (the best model-embedder combinations as detailed in Table \ref{sec:data_collection}) for five molecular properties (rows) using four regression models (columns). Blue and orange points represent training and test datasets respectively, with black lines showing ideal 1:1 correlation and orange lines showing test set linear fits. Performance metrics (R$^2$, RMSE, MAE) from 5-fold cross-validation are shown in text insets.}
  \label{fig:model_prediction}
\end{figure*}

Building upon the optimized model configurations as discussed above, we performed a detailed evaluation of prediction performance across different molecular properties using both Mol2Vec and VICGAE embeddings. Our comprehensive evaluation revealed several key insights into the effectiveness of machine learning approaches for molecular property prediction. The performance metrics across different models and properties are summarized in Table \ref{tab:model_performance} and visualized in Figure \ref{fig:r2_scores}, with several notable trends emerging from the analysis. Figure \ref{fig:model_prediction} presents a comprehensive visualization of the experimental versus predicted values across all molecular properties and regression models studied. The scatter plots reveal clear patterns in prediction accuracy, with tighter clustering around the ideal 1:1 correlation line (shown in black) for properties with higher R$^2$ values. 

\begin{table}
\caption{Performance comparison of different machine learning models using Mol2vec and VICGAE embeddings for molecular property prediction. Results are shown for melting point (MP, $^\circ$C), boiling point (BP, $^\circ$C), vapor pressure at 25$^\circ$C (VP), critical temperature (CT, K) and critical pressure (CP, MPa) of organic compounds. For each property, the best performing metrics (highest R$^2$ and lowest RMSE and MAE) are highlighted in \textbf{bold}. Values in parentheses represent the standard deviation of the last significant digit. All metrics are computed with 5-fold CV.}
\label{tab:model_performance}
\setlength{\tabcolsep}{4pt}
\scriptsize
    \begin{tabular}{llllccc}
        \toprule
        Property & Model & Embedder & R$^2$ & RMSE & MAE \\
        \midrule
        \multirow{10}{*}{MP ($^\circ$C)}\\
        & GBR & VICGAE & 0.79(2) & 43(2) & 34(1) \\
        & GBR & Mol2Vec & 0.84(2) & 39(2) & 30(2) \\
        & XGBoost & VICGAE & 0.76(1) & 46(2) & 36(1) \\
        & XGBoost & Mol2Vec & 0.82(2) & 40(2) & 31(2) \\
        & CatBoost & VICGAE & 0.79(2) & 43(1) & 34.0(9) \\
        & \textbf{CatBoost} & \textbf{Mol2Vec} & \textbf{0.86(1)} & \textbf{36(2) }& \textbf{28(1)} \\
        & LGBM & VICGAE & 0.79(1) & 43(2) & 34(1) \\
        & \textbf{LGBM} & \textbf{Mol2Vec} & \textbf{0.86(1)} & \textbf{36(2)} & \textbf{28(1)} \\
        \midrule
        \multirow{10}{*}{BP ($^\circ$C)}\\
        & GBR & VICGAE & 0.89(1) & 29(1) & 22(1) \\
        & GBR & Mol2Vec & 0.924(7) & \textbf{24(2)} & \textbf{17(1)} \\
        & XGBoost & VICGAE & 0.87(1) & 32(1) & 24(1) \\
        & XGBoost & Mol2Vec & 0.90(1) & 29(3) & 21(2) \\
        & CatBoost & VICGAE & 0.890(7) & 29.0(9) & 21.9(8) \\
        & \textbf{CatBoost} & \textbf{Mol2Vec} & \textbf{0.925(8)} & \textbf{24(2)} & \textbf{17(1)} \\
        & LGBM & VICGAE & 0.88(1) & 30(1) & 23(1) \\
        & LGBM & Mol2Vec & 0.92(1) & 26(2) & 18(1) \\
        \midrule
        \multirow{10}{*}{VP (KPa)}\\
        & \textbf{GBR} & \textbf{VICGAE} & \textbf{0.4(2)} &\textbf{ 2.1(5)} & \textbf{1.5(3)} \\
        & GBR & Mol2Vec & 0.31(1) & 2.3(1) & 1.7(1) \\
        & XGBoost & VICGAE & 0.3(2) & 2.3(7) & 1.6(4) \\
        & XGBoost & Mol2Vec & 0.2(2) & 2.5(3) & 1.8(2) \\
        & \textbf{CatBoost} & \textbf{VICGAE} & \textbf{0.4(2)} & \textbf{2.1(5)} & \textbf{1.5(3)} \\
        & CatBoost & Mol2Vec & 0.32(7) & 2.3(2) & 1.7(1) \\
        & \textbf{LGBM} & \textbf{VICGAE} & \textbf{0.4(2) }& \textbf{2.1(5) }& \textbf{1.5(3)} \\
        & LGBM & Mol2Vec & 0.3(1) & 2.32(8) & 1.73(6) \\
        \midrule
        \multirow{10}{*}{CT (K)}\\
        & GBR & VICGAE & 0.90(2) & 37(4) & 28(3) \\
        & GBR & Mol2Vec & 0.93(1) & 33(4) & 21(3) \\
        & XGBoost & VICGAE & 0.88(2) & 41(4) & 32(3) \\
        & XGBoost & Mol2Vec & 0.90(1) & 39(4) & 28(3) \\
        & CatBoost & VICGAE & 0.90(2) & 38(4) & 29(3) \\
        & \textbf{CatBoost} & \textbf{Mol2Vec} & \textbf{0.931(7)} & \textbf{32(3)} & \textbf{20(3)} \\
        & LGBM & VICGAE & 0.89(2) & 38(3) & 29(2) \\
        & LGBM & Mol2Vec & 0.91(1) & 36(3) & 24(3) \\
        \midrule
        \multirow{10}{*}{CP (MPa)}\\
        & GBR & VICGAE & 0.90(1) & 0.34(3) & 0.25(2) \\
        & GBR & Mol2Vec & 0.91(2) & 0.32(4) & 0.21(3) \\
        & XGBoost & VICGAE & 0.88(2) & 0.38(3) & 0.28(2) \\
        & XGBoost & Mol2Vec & 0.89(1) & 0.35(4) & 0.24(2) \\
        & CatBoost & VICGAE & 0.90(2) & 0.35(2) & 0.25(2) \\
        & \textbf{CatBoost} & \textbf{Mol2Vec} & \textbf{0.92(2)} & \textbf{0.31(4)} & \textbf{0.19(2)} \\
        & LGBM & VICGAE & 0.89(1) & 0.38(3) & 0.27(2) \\
        & LGBM & Mol2Vec & 0.91(2) & 0.33(5) & 0.22(2) \\
        \bottomrule
    \end{tabular}
\end{table}

As depicted in Table \ref{tab:model_performance}, Critical Temperature (CT) and Critical Pressure (CP) demonstrated excellent prediction performance, with CatBoost using Mol2Vec achieving R$^2$ scores of 0.931(7) and 0.92(2) respectively. The performance can be attributed to well-balanced datasets and moderate distribution characteristics. Our critical temperature and pressure models (R² $\approx 0.93$) exhibit excellent accuracy, aligning with or exceeding previous QSPR models \cite{banchero_comparison_2018}.

Boiling Point (BP) predictions also showed strong performance with CatBoost and GBR achieving R$^2$ values of 0.925(8) using Mol2Vec embeddings. The large dataset size likely contributed to this robust performance.

Melting Point (MP) predictions achieved relatively moderate performance with CatBoost and LGBM using Mol2Vec embeddings (R$^2$ = 0.86(1), RMSE = 36(2)$^\circ$C), despite having the largest dataset. This accuracy is comparable to the best literature reports for large MP datasets. In previous studies, ML models achieved MP errors on the order of $30-40 ^\circ$C \cite{tetko_how_2014, mcdonagh_predicting_2015}, so our result of RMSE $\approx 36 ^\circ$C) is on par with state-of-the-art predictions.

The relationship between prediction accuracy and data characteristics becomes evident when examining the distribution patterns of target properties (see Figures \ref{fig:y_distribution}, \ref{fig:r2_scores}, and \ref{fig:model_prediction}). Additionally, it is also evident from the residual plot as shown in Figure S2 (Supporting
Information). CT shows nearly normal distribution (skewness: 0.0014) and corresponds to the highest model performance. BP and MP, with slight skewness (0.13 and 0.081 respectively), also achieve strong predictions. 

In contrast, vapor pressure predictions proved the most challenging, corresponding to the highest skewness (1.1) and smallest dataset size. Notably, the limited VP prediction performance also reflects the inability of the current embeddings to explicitly capture non-covalent intermolecular interactions, such as hydrogen bonding or dipole effects, which are essential for accurate modeling of vapor pressure.

Comparing the performance between Mol2Vec and VICGAE embeddings, the accuracy differences vary across properties:

\begin{itemize}
\item For MP prediction: 9(1)\% difference (R$^2$ 0.86(1) vs 0.79(2))
\item For BP prediction: 4(1)\% difference (R$^2$ 0.925(8) vs 0.890(7))
\end{itemize}

\begin{figure}[!htb]
  \includegraphics[width=\columnwidth,keepaspectratio]{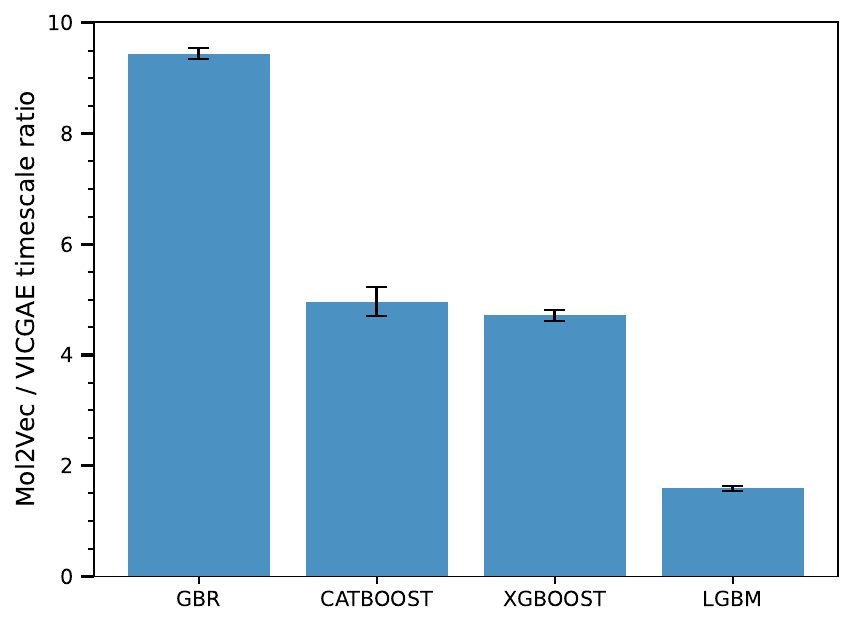}
  \caption{Comparison of computational execution time between Mol2Vec and VICGAE embeddings across different regression models for MP dataset. Benchmarks were conducted on an Apple M2 Pro processor (10-Core CPU, 32 GB RAM). The black solid lines within the bars represent error bars, indicating the uncertainty in the measured speedup factors. VICGAE demonstrates notable computational efficiency gains across all regression models.}
  \label{fig:execution_timescales}
\end{figure}

VICGAE demonstrated comparable performance characteristics across different molecular properties. For VP prediction, while VICGAE showed nominally higher mean R² values 0.4(2) compared to Mol2Vec 0.32(7), the overlapping uncertainty ranges suggest that the performance difference may not be statistically significant. Notably, VICGAE achieved this competitive performance despite its significantly lower dimensionality (32 dimensions vs Mol2Vec's 300 dimensions). Beyond accuracy, VICGAE showed remarkable computational efficiency. As shown in Figure \ref{fig:execution_timescales}, for the MP property (our largest dataset), VICGAE consistently achieved faster execution times across all models, with GBR showing approximately 10x speedup compared to Mol2Vec. These efficiency gains, combined with competitive performance even for challenging properties like VP prediction, make VICGAE particularly attractive for high-throughput screening applications where both speed and accuracy are crucial.

\section{Conclusion}

This comprehensive study evaluated the performance of various machine learning models combined with two molecular embedding approaches (Mol2vec and VICGAE) for predicting five crucial molecular properties. Our findings demonstrate that model performance is influenced by both the underlying data characteristics and the chosen embedding method, revealing several key insights.

\textbf{Key Findings}: The study demonstrated strong prediction capabilities across multiple properties, achieving R$^2$ values up to 0.93 for critical temperature and 0.925 for boiling point predictions using CatBoost with Mol2Vec embeddings. VICGAE showed R$^2$ values of 0.4(2) compared to Mol2Vec's 0.32(7), with overlapping uncertainty ranges suggesting comparable performance despite having the smallest dataset. This highlights the potential advantage of lower-dimensional embeddings for maintaining competitive performance in challenging prediction tasks with limited data availability.

\textbf{Embedding Method Performance}: A significant finding emerged from the comparison of embedding techniques. While Mol2Vec (300 dimensions) showed superior prediction accuracy for most properties, VICGAE's 32-dimensional embeddings demonstrated competitive performance and even outperformed Mol2Vec for vapor pressure prediction. The observed 10x improvement in execution speed, combined with superior performance for challenging properties, establishes VICGAE as a particularly promising option for high-throughput molecular screening applications. This balance of efficiency and accuracy provides valuable flexibility for various applications in computational chemistry, especially when dealing with limited datasets or computationally intensive screening tasks.

\textbf{Model Optimization Insights}: Our detailed hyperparameter optimization analysis revealed that while Gradient Boosting Regression benefits substantially from careful tuning, modern frameworks like CatBoost, LightGBM, and XGBoost demonstrate robust performance even with default configurations. This finding has practical implications for reducing computational overhead in model development while maintaining prediction accuracy.

\textbf{Practical Implications}: The insights gained from this study contribute to the broader field of computer-aided molecular design by providing clear guidelines for selecting appropriate embedding methods and machine learning models based on specific application requirements. The demonstrated success in predicting various molecular properties suggests that these approaches can significantly accelerate the molecular discovery process while reducing experimental costs. Importantly, ChemXploreML’s user-friendly design makes ML modeling accessible to chemists with limited programming expertise, bridging the gap between computational and experimental workflows.

\textbf{ChemXploreML}: The desktop application developed in this study, with its flexible architecture and efficient implementation, provides a solid foundation for future developments in molecular property prediction. As the field continues to evolve, the integration of more sophisticated embedding techniques and advanced machine learning architectures will likely further improve prediction capabilities, particularly for challenging properties with limited experimental data.

\textbf{Future Perspectives}: This study opens several promising avenues for ongoing research in molecular property prediction. First, there is a clear need to develop specialized embedding architectures that can better handle highly skewed property distributions, particularly for challenging cases like vapor pressure prediction. The success of VICGAE's compact embeddings suggests potential for architectural innovations that could further optimize the trade-off between dimensionality and predictive power. Investigation of hybrid embedding approaches could leverage the complementary strengths of different techniques while maintaining computational efficiency. These directions, collectively, would advance the field toward more reliable and efficient computational tools for molecular design and discovery.

In parallel, the implementation of applicability domain (AD) analysis—via leverage score and Mahalanobis distance methods—within ChemXploreML provides a foundation for assessing the reliability of model predictions. Although not explored in this study, this feature enables identification of out-of-distribution compounds and will be instrumental in future efforts aimed at high-confidence molecular screening. These directions, collectively, would advance the field toward more reliable and efficient computational tools for molecular design and discovery.

\section{Data and Software \text{Availability}}
All data utilized in this study, including molecular property datasets (original, validated for each embedder, and cleaned), molecular embeddings, and the model configuration file, are provided in a publicly accessible Supporting Information file available at \url{https://zenodo.org/doi/10.5281/zenodo.15007626}. The data can be accessed through the ChemXploreML desktop application, which was specifically developed in this study for molecular property prediction, and is publicly available via Zenodo at \citet{aravindh_nivas_marimuthu_aravindhnivaschemxploreml_2025}. The latest version and releases of ChemXploreML are available at \url{https://github.com/aravindhnivas/ChemXploreML/releases}.

Installation instructions for ChemXploreML, including platform-specific notes and download links, are provided in Section 4 of the Supporting Information. Please note that macOS builds are currently unsigned and may trigger a security warning during installation.

\section{Author Contributions}
Aravindh N. Marimuthu developed the ChemXploreML desktop application, implemented all machine learning pipelines, performed data preprocessing and analysis, and wrote the initial draft of the manuscript. Brett A. McGuire supervised the project, provided conceptual guidance on molecular property prediction and manuscript structure. 

The authors declare no competing financial interests.

\section{Acknowledgment}
This work was supported by the Schmidt Family Foundation. The authors thank the Massachusetts Institute of Technology Department of Chemistry for additional institutional support and access to computational resources.

\begin{suppinfo}

Additional figures and details such as UMAP visualizations with perceptually uniform colormaps; residual and joint distribution plots for molecular properties; clustering analysis with functional group, ring-type, and structure-type breakdowns across MP, BP, VP, CT, and CP datasets are provided in the Supporting Information (PDF).

\end{suppinfo}

\bibliography{references, softwares, github-tools}

\end{document}


\section{Abbreviations}

\begin{itemize}
    \item MP: Melting Point ($^\circ$C)
    \item BP: Boiling Point  ($^\circ$C)
    \item VP: Vapor Pressure (KPa at 25 $^\circ$C)
    \item CP: Critical Pressure (MPa)
    \item CT: Critical Temperature (K)
    \item MW: Molecular Weight (u)
    \item RMSE: Root Mean Square Error (property specific units) 
    \item MAE: Mean Absolute Error (property specific units) 
    \item R$^2$: Coefficient of determination (dimensionless)
    \item GBR: Gradient boosting Regressor
    \item LGBM: Light Gradient Boosting Machine
    \item CV: Cross validation
\end{itemize}

\section{Statistics}

\subsection{Metrics}
\label{sec:SI_metrics}

The performance of the models was evaluated using three standard statistical metrics:
\begin{equation}
    RMSE = \sqrt{\frac{1}{n}\sum_{i=1}^{n}(y_i - \hat{y}_i)^2}
\end{equation}
\begin{equation}
    MAE = \frac{1}{n}\sum_{i=1}^{n}|y_i - \hat{y}_i|
\end{equation}
\begin{equation}
    R^2 = 1 - \frac{\sum_{i=1}^{n}(y_i - \hat{y}i)^2}{\sum{i=1}^{n}(y_i - \bar{y})^2}
\end{equation}
where $y_i$ and $\hat{y}_i$ are the experimental and predicted values respectively, $\bar{y}$ is the mean of experimental values, and $n$ is the number of samples. RMSE and MAE provide error measurements in the original units of the property, while $R^2$ indicates the proportion of variance explained by the model.

\subsection{N-Fold Cross Validation}
\label{sec:SI_n_fold}

In N-fold cross-validation, the dataset is partitioned into N equal-sized subsets. The model is trained on N-1 subsets and validated on the remaining subset. This process is repeated N times, with each subset serving as the validation set exactly once. The final performance metrics are averaged across all N iterations.
\begin{equation}
CV_{score} = \frac{1}{N}\sum_{i=1}^{N} Score_i
\end{equation}
where $Score_i$ represents the performance metric (RMSE, MAE, or $R^2$) for the i-th fold. In this work, we primarily used 5-fold cross-validation (N=5) to assess model generalization performance. The standard deviation of these N scores provides an estimate of the model's robustness:
\begin{equation}
\sigma = \sqrt{\frac{1}{N-1}\sum_{i=1}^{N}(Score_i - \overline{Score})^2}
\end{equation}
where $\overline{Score}$ is the mean score across all folds. Results are reported as $CV_{score} \pm \sigma$ to indicate both the average performance and its variation.

\begin{figure*}[!htb]
  \includegraphics[width=1\textwidth,keepaspectratio]{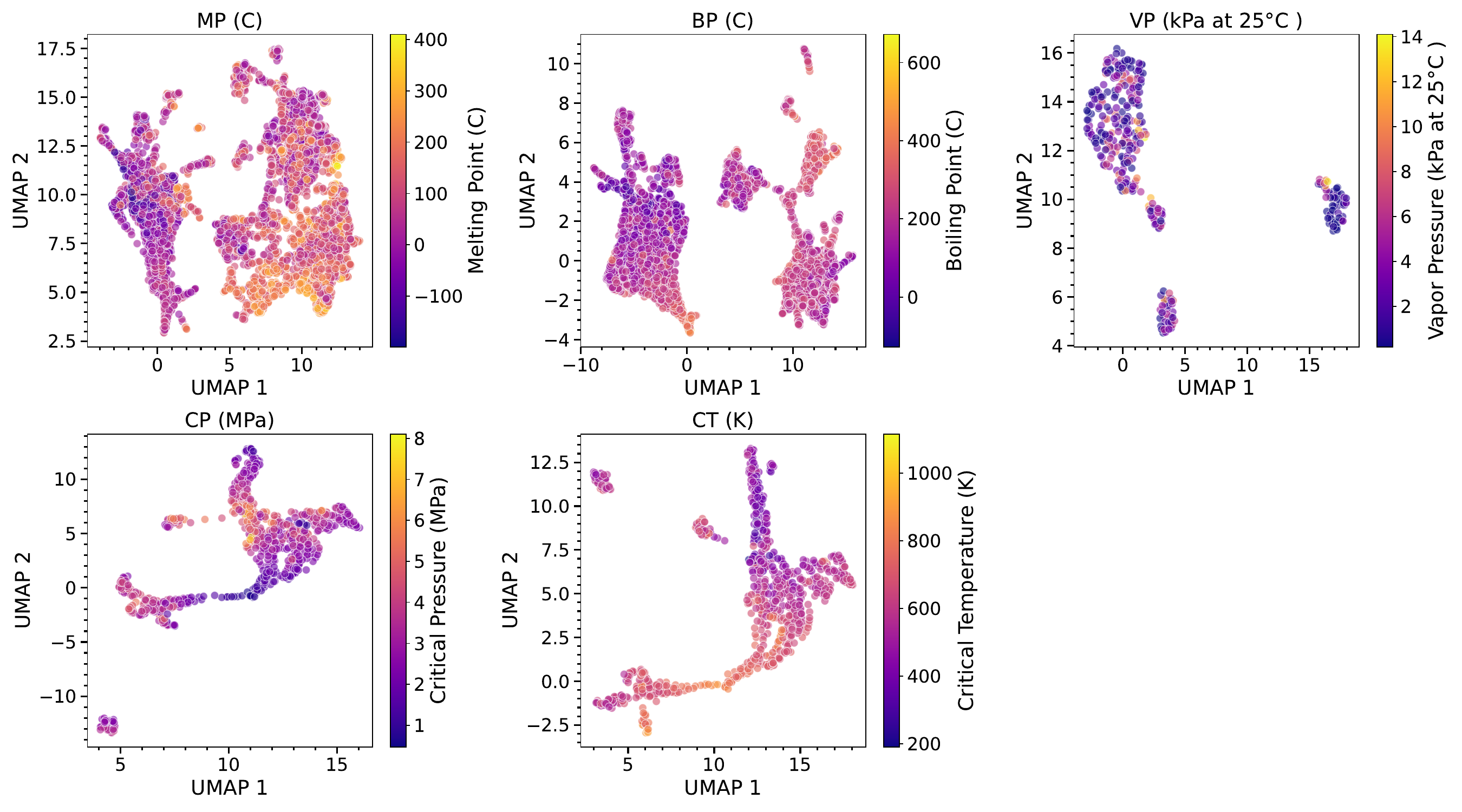}
  \caption{UMAP projections illustrating the clustering patterns of molecular embeddings for five distinct properties: melting point (MP, $^{\circ}$C), boiling point (BP, $^{\circ}$C), vapor pressure (VP, kPa at 25$^{\circ}$C), critical pressure (CP, MPa), and critical temperature (CT, K). Each subplot presents a two-dimensional representation of the high-dimensional molecular data, with points colored according to their respective property values. 
  }
  \label{fig:umap_perceptually_uniform}
\end{figure*}

\begin{figure}[!htb]
  \includegraphics[width=\columnwidth,keepaspectratio]{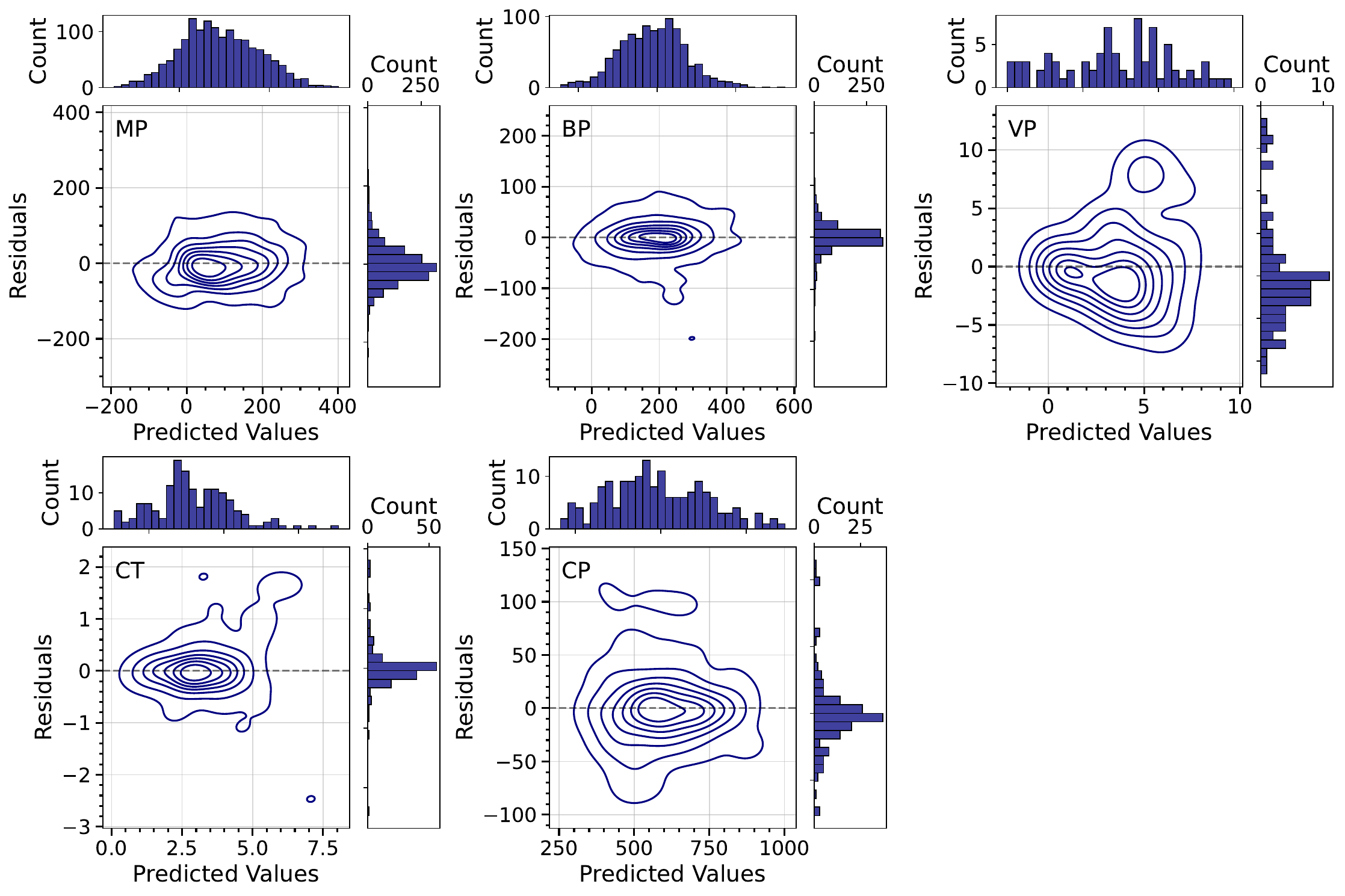}
  \caption{Residual analysis plots for molecular property predictions using CatBoost with Mol2Vec embeddings. The plots show predicted values versus residuals for five molecular properties: melting point (MP, $^\circ$C), boiling point (BP, $^\circ$C), vapor pressure (VP, kPa at 25 $^\circ$C), critical temperature (CT, K), and critical pressure (CP, MPa). Contour plots (center) indicate prediction density distributions, while marginal histograms show the distribution of predicted values (top) and residuals (right). The symmetric patterns and normal distributions in CT and BP indicate robust model performance, while the heteroscedastic pattern in VP suggests higher prediction uncertainty. Dashed lines represent zero residual reference.}
  \label{fig:SI_residual_plot}
\end{figure}

\begin{figure}[!htb]
  \includegraphics[width=\columnwidth,keepaspectratio]{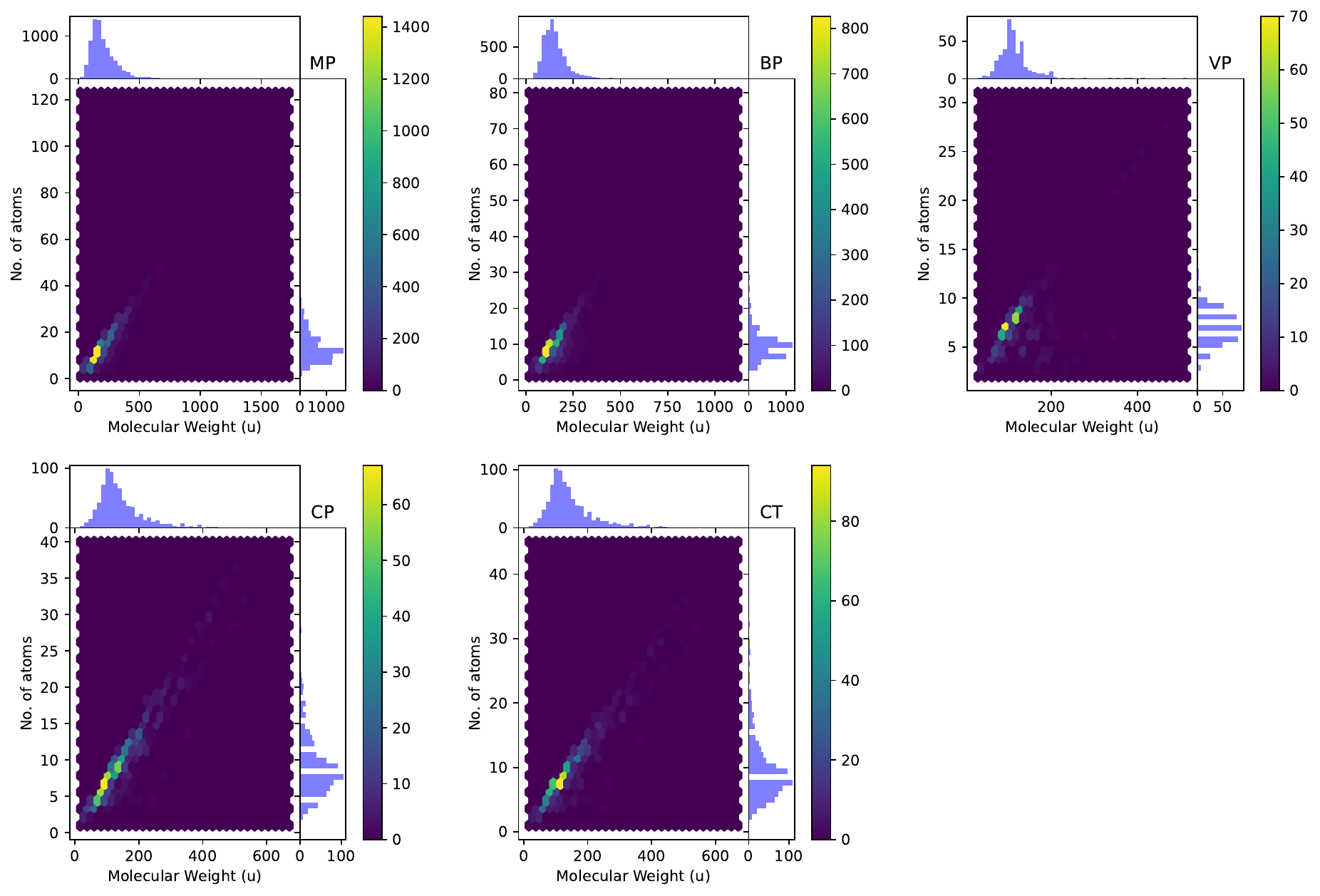}
  \caption{Joint distribution analysis of molecular weight and number of atoms across five molecular properties (MP, BP, VP, CP, and CT) datasets.}
  \label{fig:SI_mw_vs_atomic_size_distribution}
\end{figure}

\clearpage

\section{Cluster Analysis Figures}
\label{SA-sec:all-figs}

\subsection{MP}

\begin{figure*}[!htb]
    \centering
    \begin{subfigure}[b]{0.7\linewidth}
        \centering
        \includegraphics[width=\linewidth]{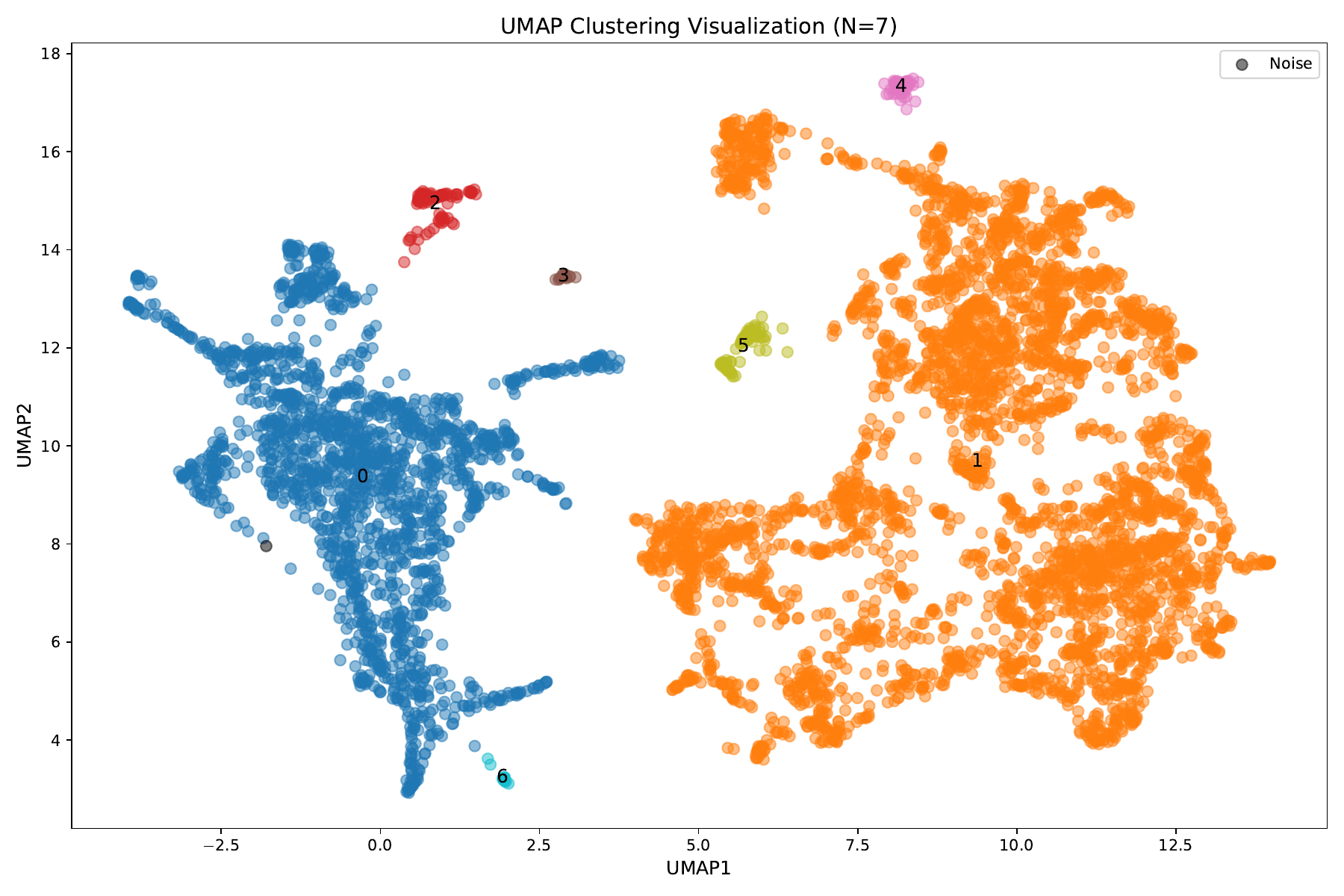}
        \caption{}
    \end{subfigure}
    \begin{subfigure}[b]{0.7\linewidth}
        \centering
        \includegraphics[width=\linewidth]{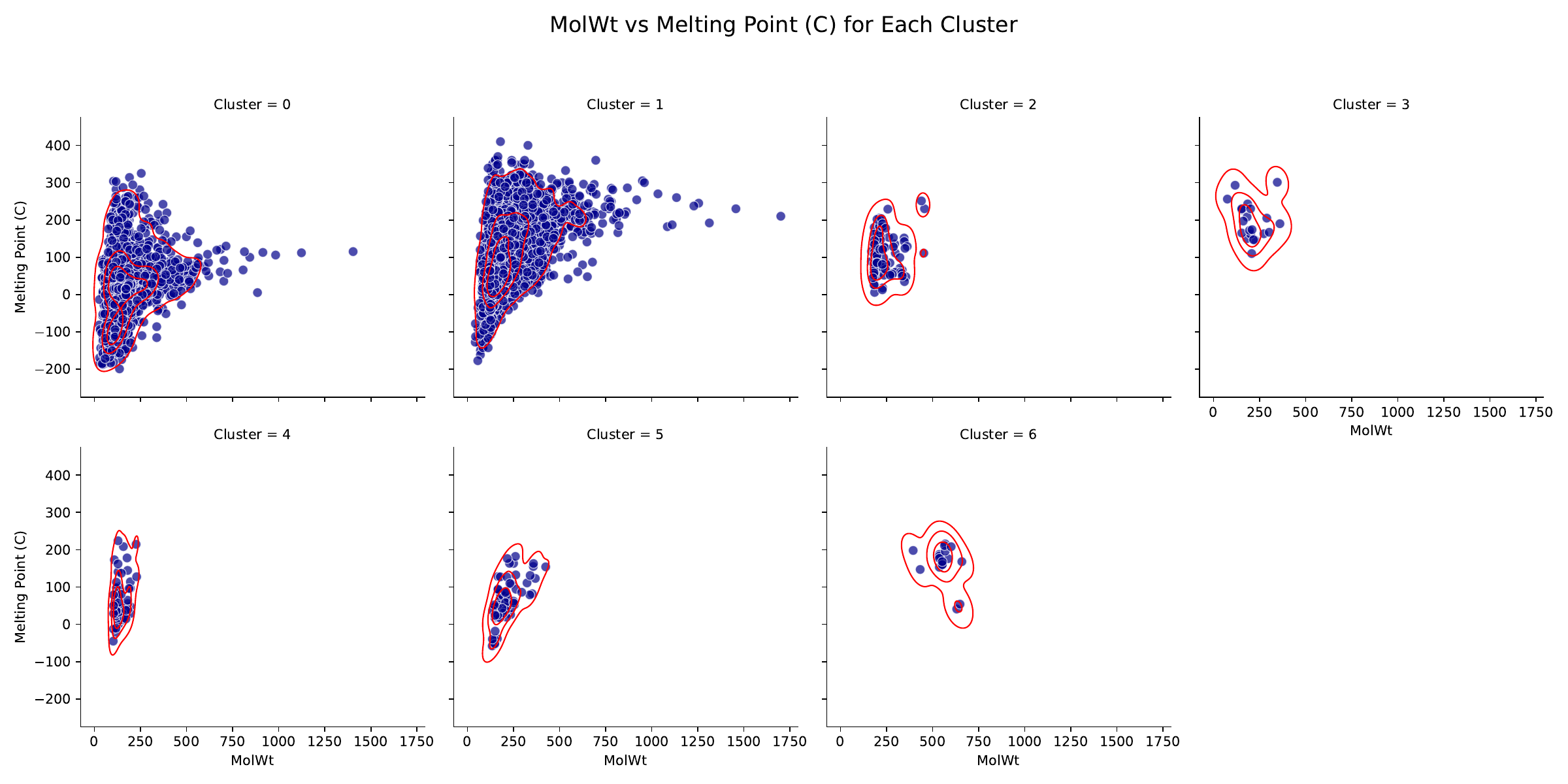}
        \caption{)}
    \end{subfigure}
    \caption{MP Cluster analysis: (a) UMAP representation of molecular embeddings, color-coded by identified clusters. (b) Relationship between molecular weight (MolWt, u) and MP ($^\circ$C).}
\end{figure*}

\begin{figure*}[!htb]
    \centering
    \begin{subfigure}[b]{\linewidth}
        \centering
        \includegraphics[width=\linewidth]{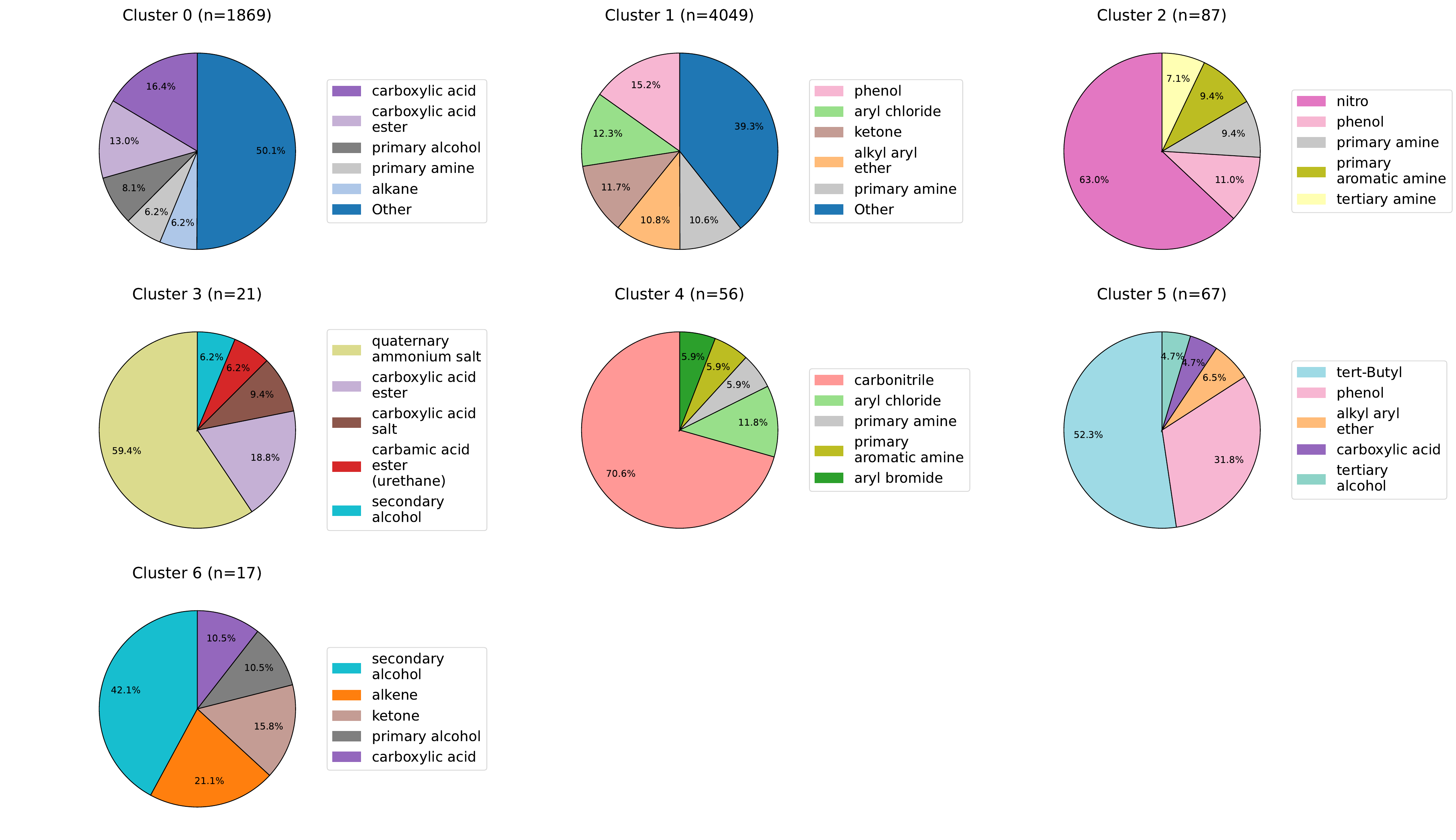}
        \caption{}
    \end{subfigure}
    \begin{subfigure}[b]{\linewidth}
        \centering
        \includegraphics[width=\linewidth]{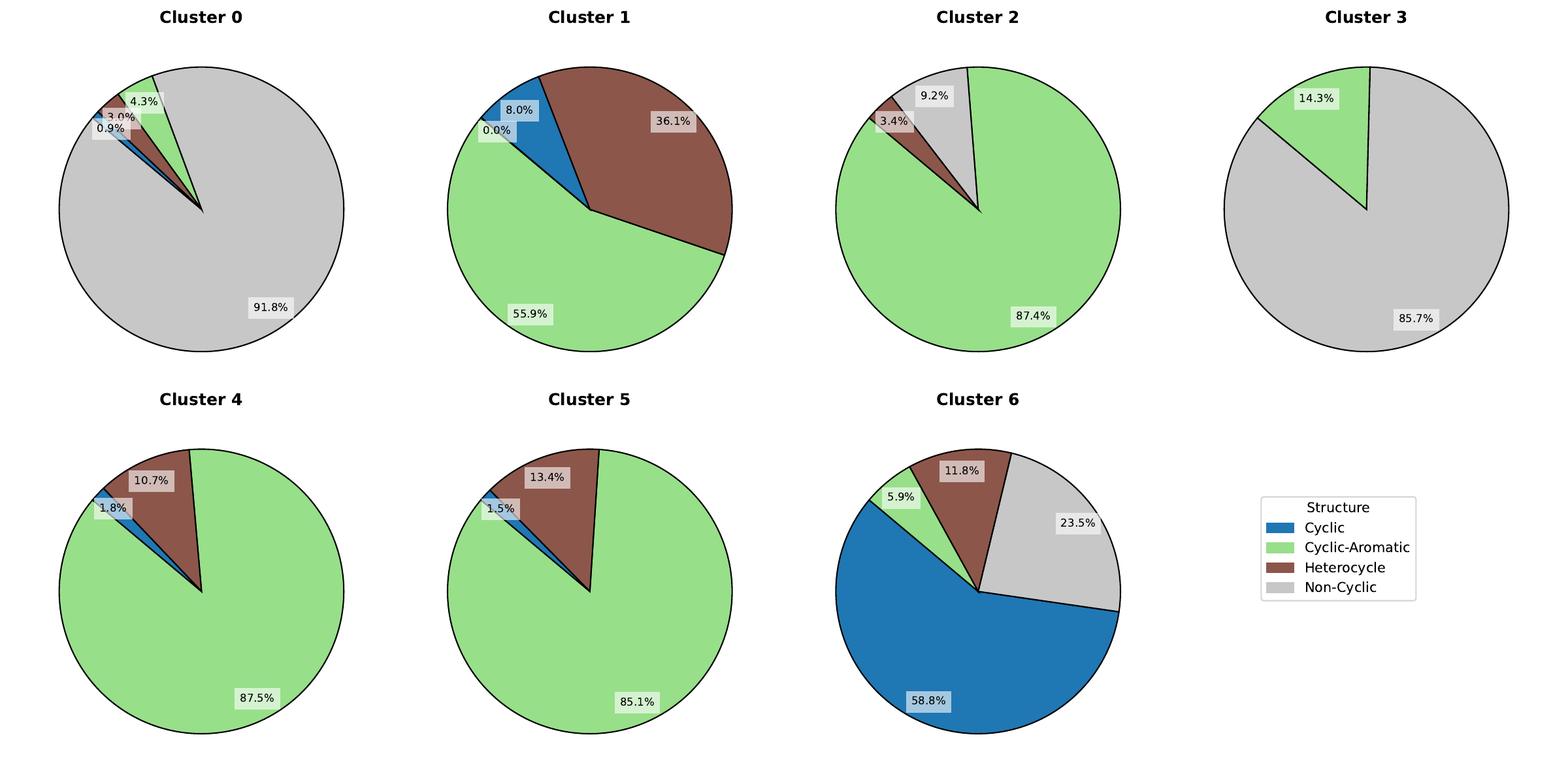}
        \caption{}
    \end{subfigure}
    \caption{MP Cluster analysis: (a) Functional group distribution across clusters. (b) Structural composition analysis shows the proportion of cyclic, aromatic, and non-cyclic compounds per cluster, with cyclic and highly conjugated systems favouring higher MPs.}
\end{figure*}

\begin{figure*}[!htb]
    \centering
     \begin{subfigure}[b]{\linewidth}
        \centering
        \includegraphics[width=\linewidth]{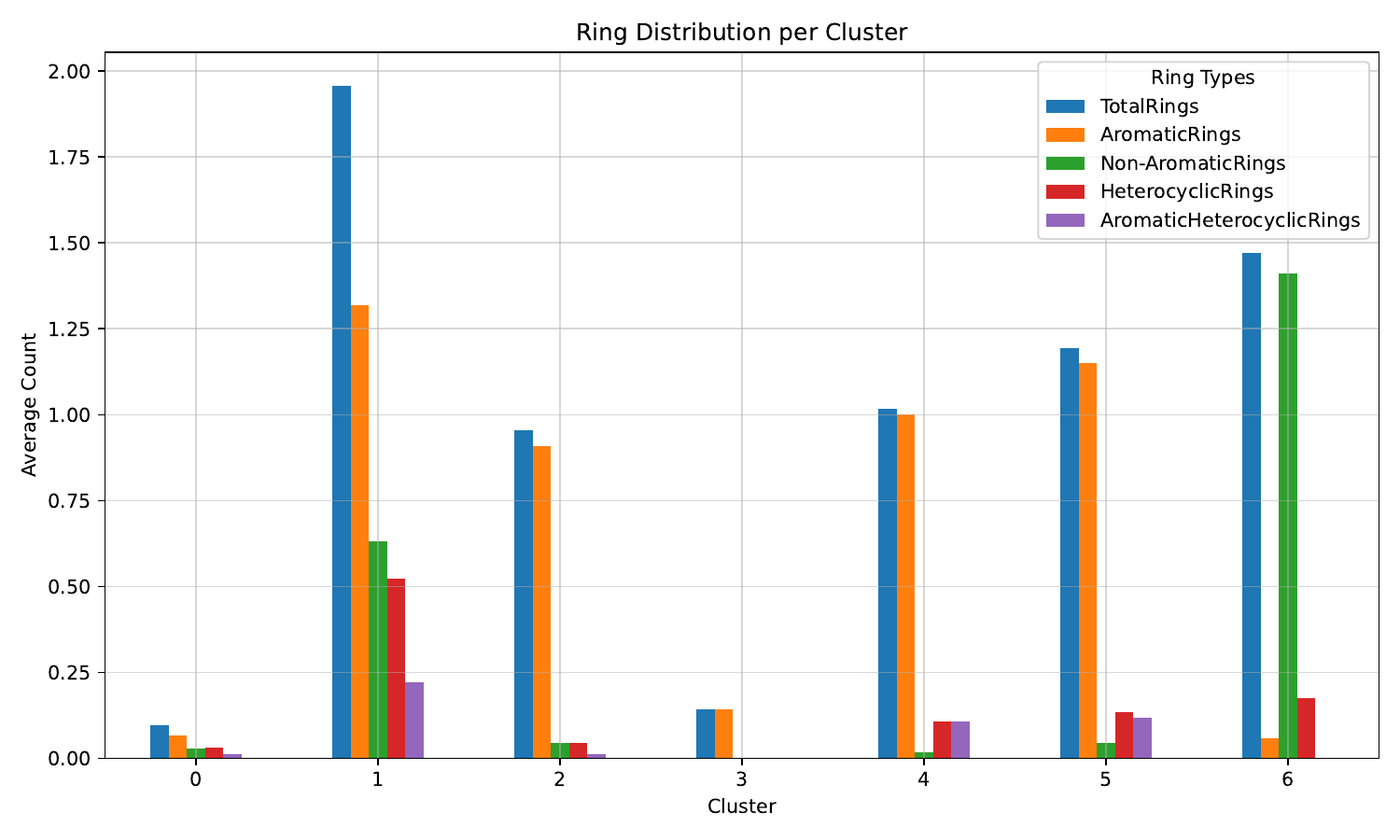}
        \caption{}
    \end{subfigure}
    \begin{subfigure}[b]{\linewidth}
        \centering
        \includegraphics[width=\linewidth]{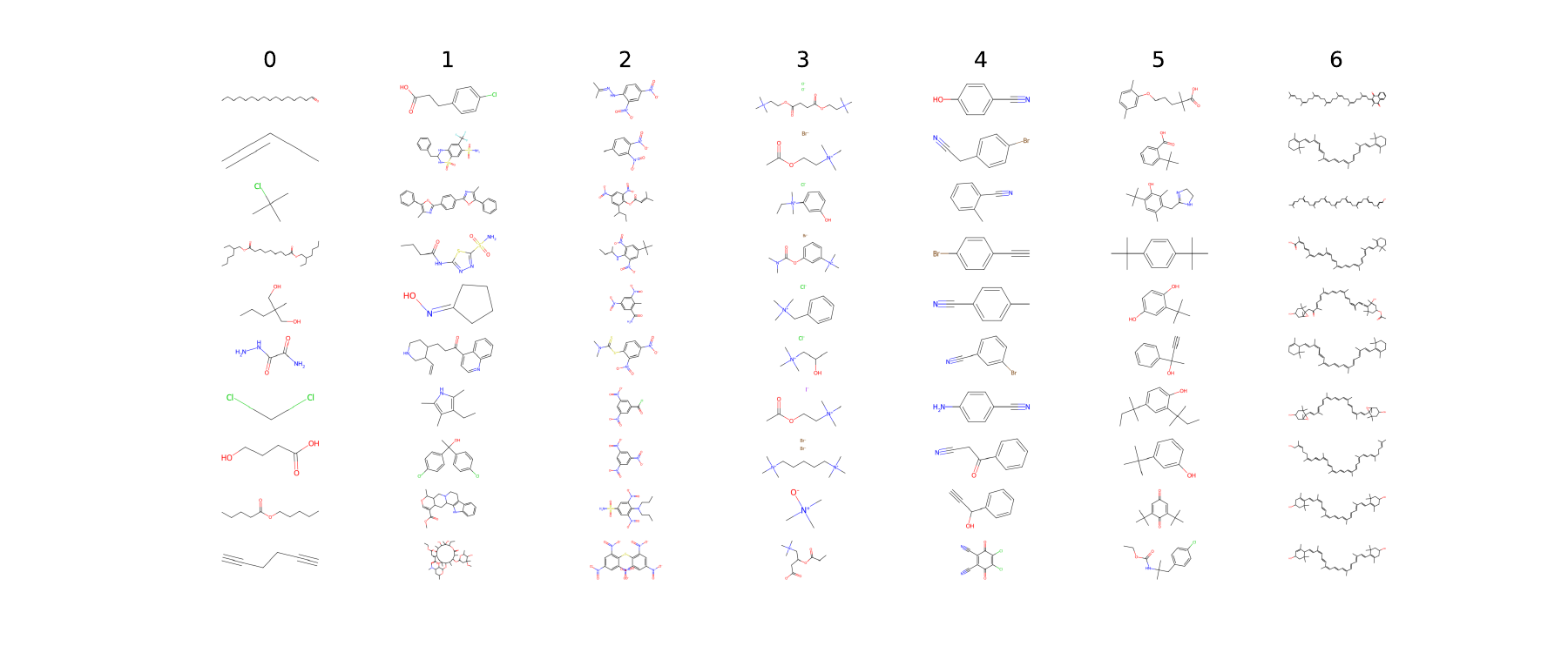}
        \caption{}
    \end{subfigure}
    \caption{MP Cluster analysis: (a) Average ring count per cluster for different ring types. (b) Representative molecular structures from each cluster}
\end{figure*}

\clearpage

\subsection{BP}

\begin{figure*}[!htb]
    \centering
    \begin{subfigure}[b]{0.7\linewidth}
        \centering
        \includegraphics[width=\linewidth]{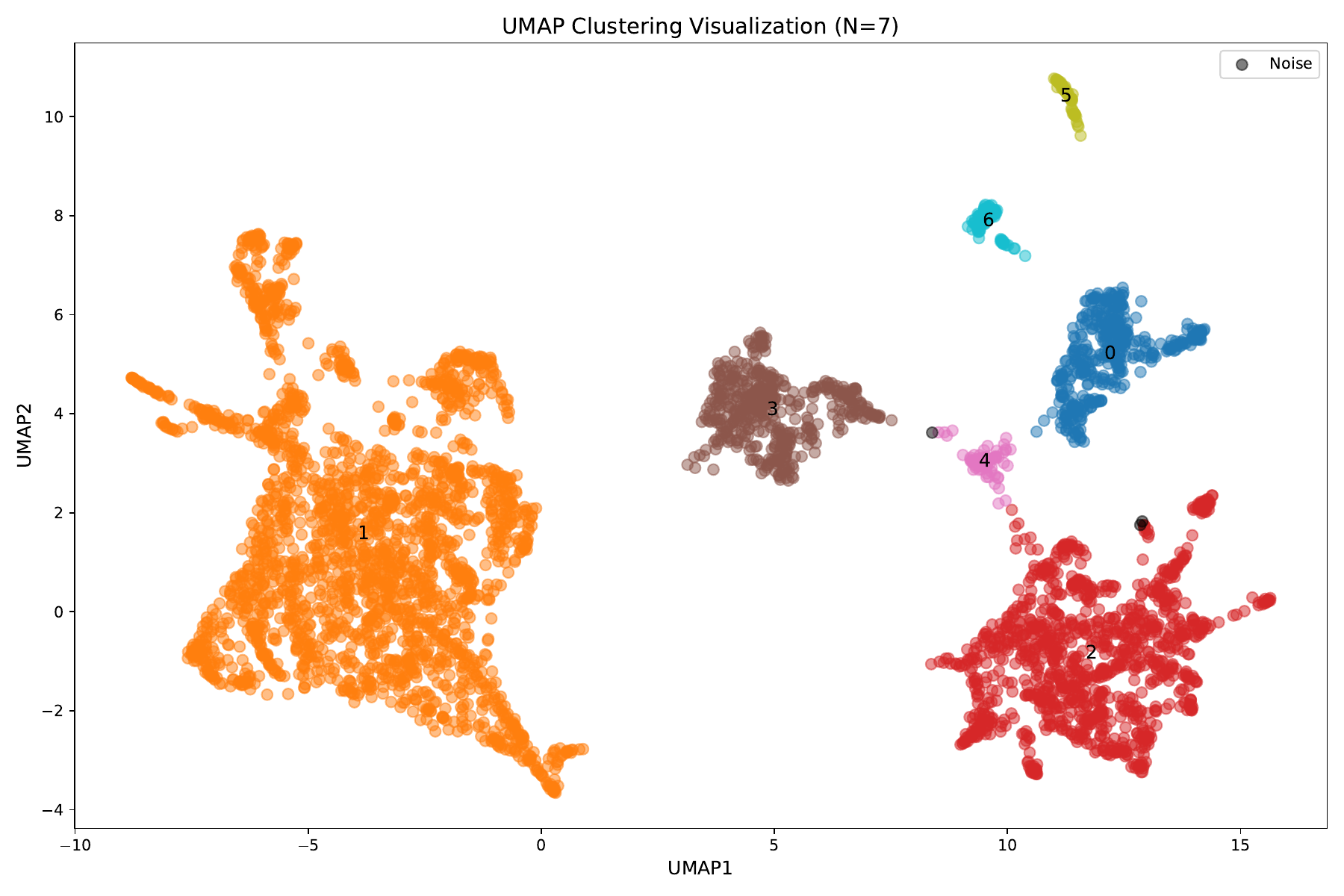}
        \caption{}
    \end{subfigure}
    \begin{subfigure}[b]{0.7\linewidth}
        \centering
        \includegraphics[width=\linewidth]{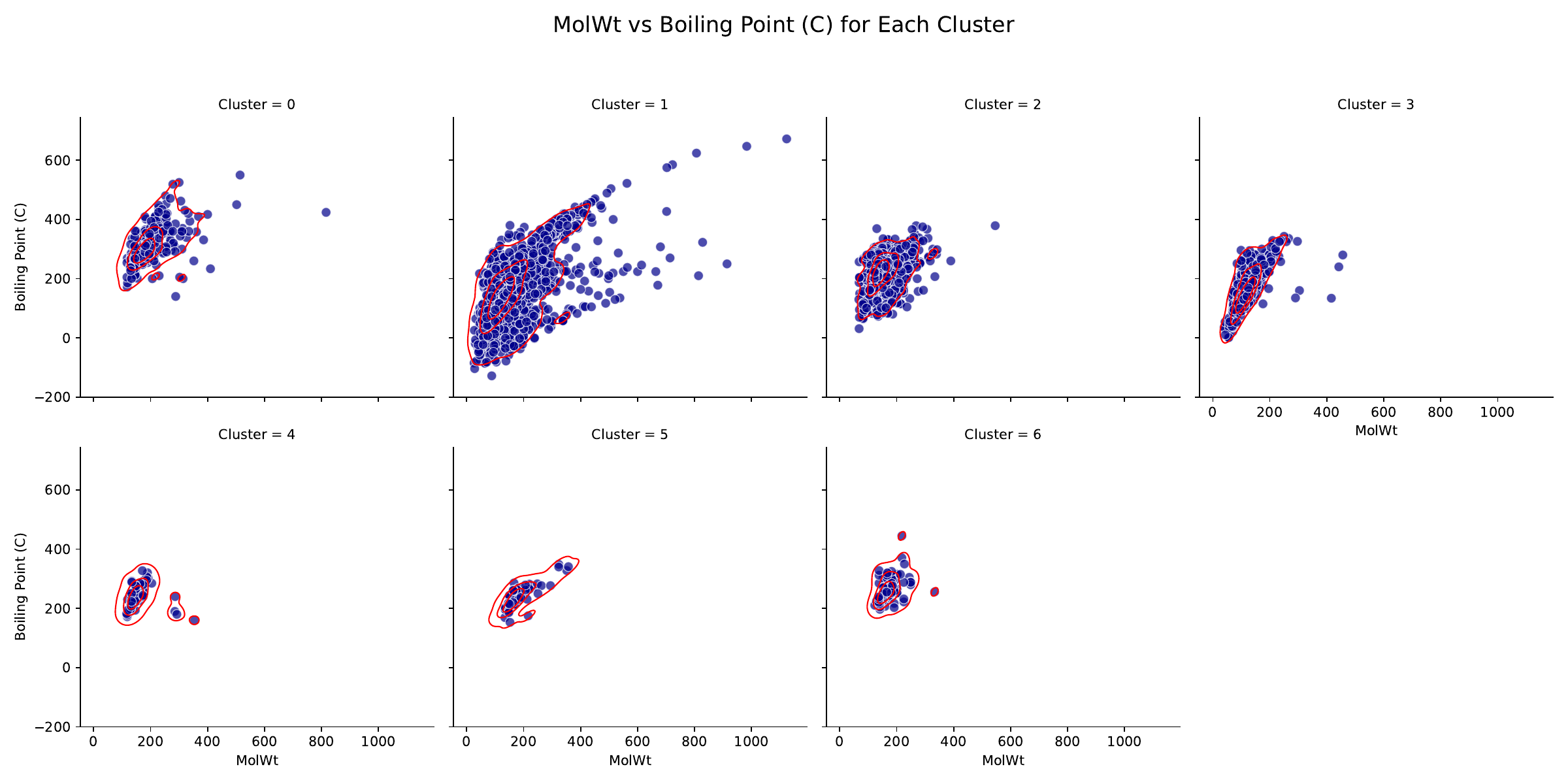}
        \caption{)}
    \end{subfigure}
    \caption{BP Cluster analysis: (a) UMAP representation of molecular embeddings, color-coded by identified clusters. (b) Relationship between molecular weight (MolWt, u) and BP ($^\circ$C).}
\end{figure*}

\begin{figure*}[!htb]
    \centering
    \begin{subfigure}[b]{\linewidth}
        \centering
        \includegraphics[width=\linewidth]{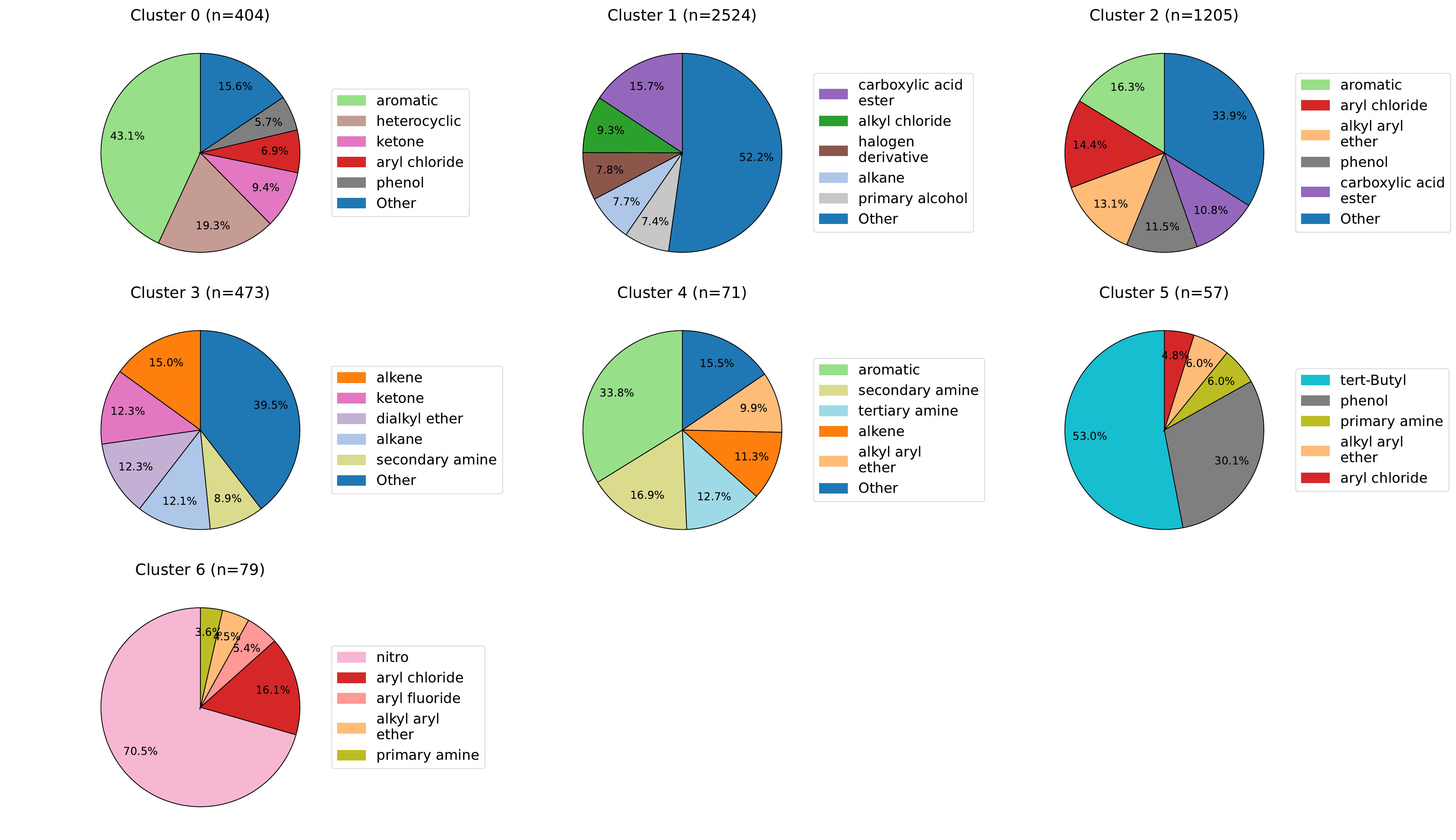}
        \caption{}
    \end{subfigure}
    \begin{subfigure}[b]{\linewidth}
        \centering
        \includegraphics[width=\linewidth]{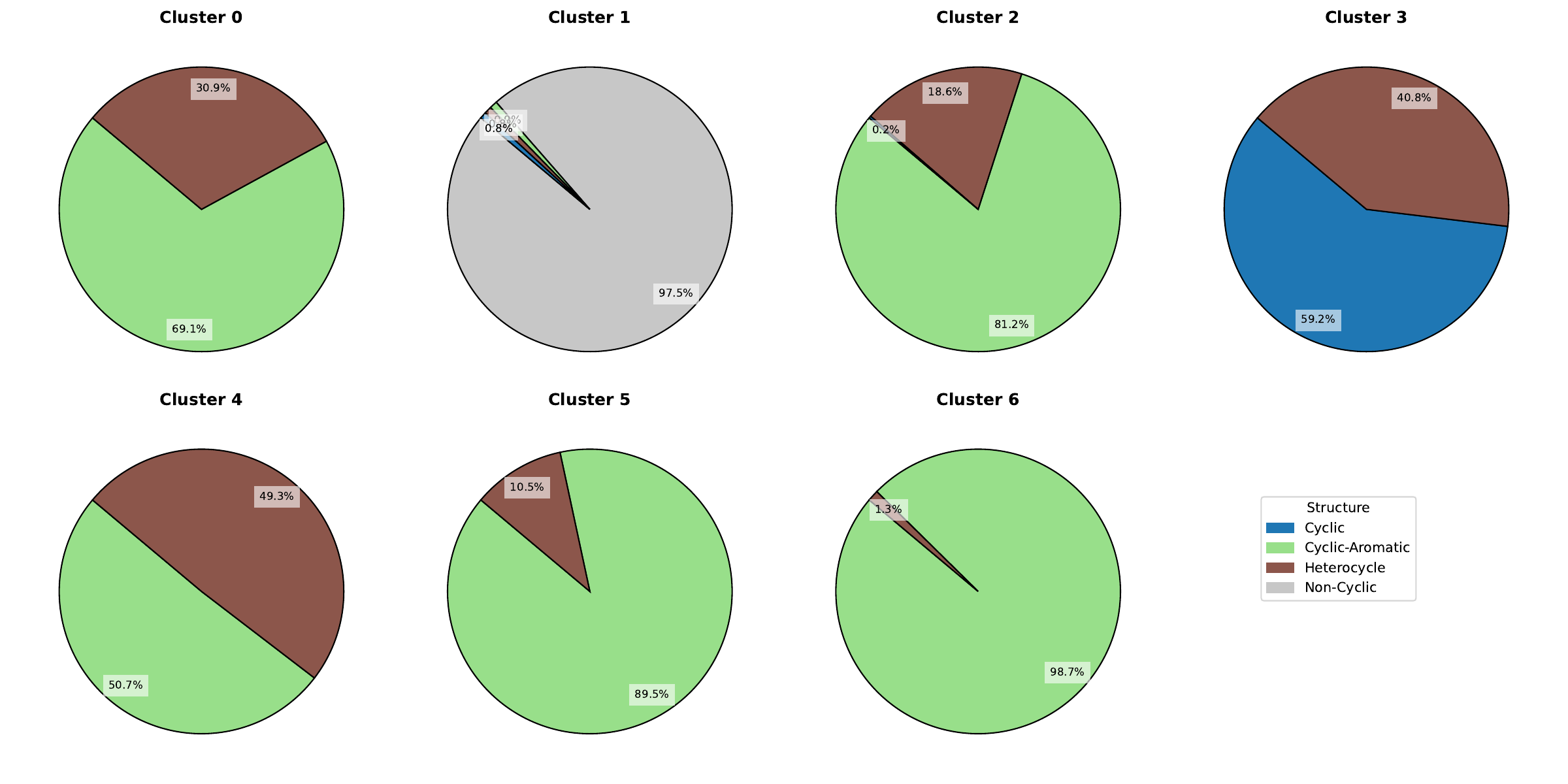}
        \caption{}
    \end{subfigure}
    \caption{BP Cluster analysis: (a) Functional group distribution across clusters. (b) Structural composition analysis shows the proportion of cyclic, aromatic, and non-cyclic compounds per cluster, with cyclic and highly conjugated systems favouring higher MPs.}
\end{figure*}

\begin{figure*}[!htb]
    \centering
     \begin{subfigure}[b]{\linewidth}
        \centering
        \includegraphics[width=\linewidth]{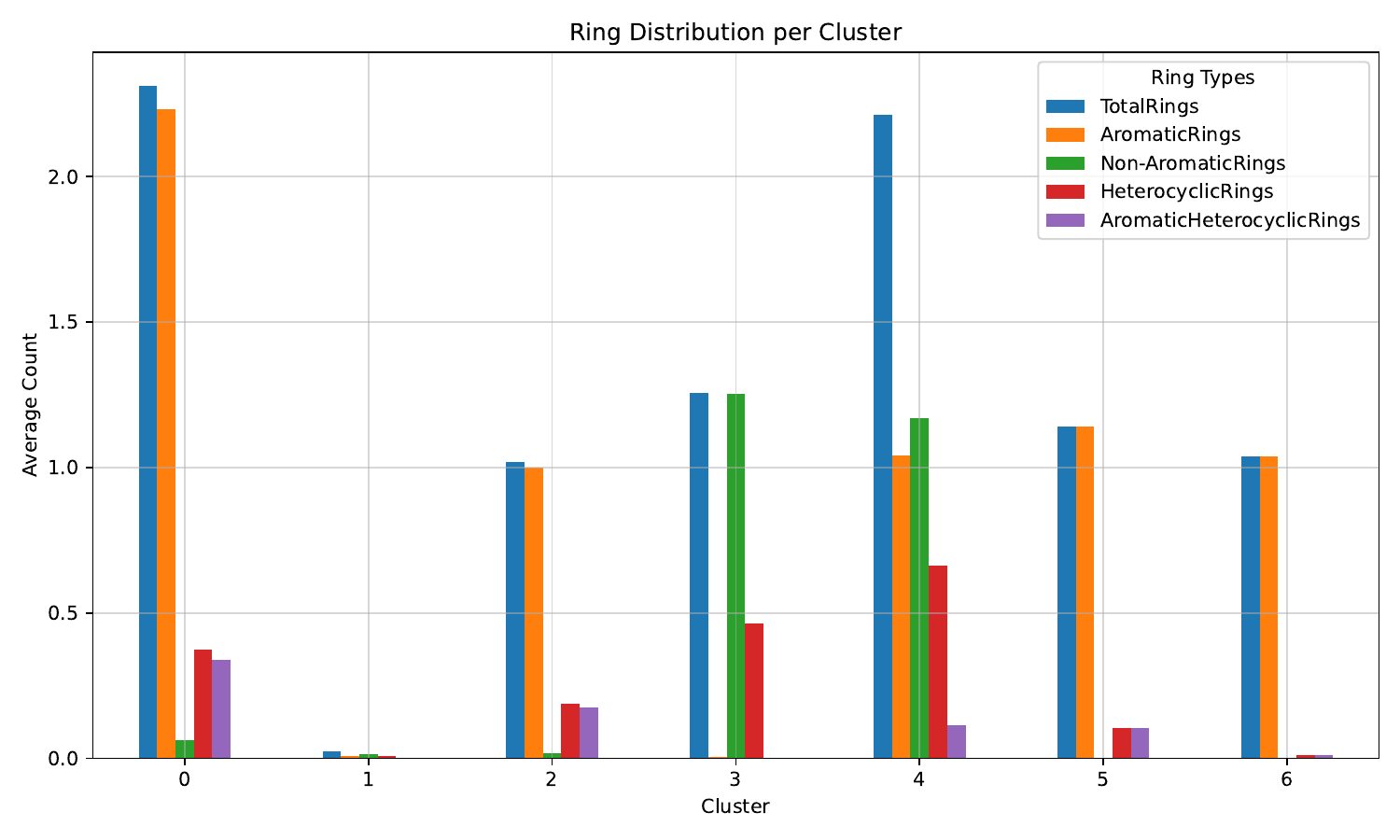}
        \caption{}
    \end{subfigure}
    \begin{subfigure}[b]{\linewidth}
        \centering
        \includegraphics[width=\linewidth]{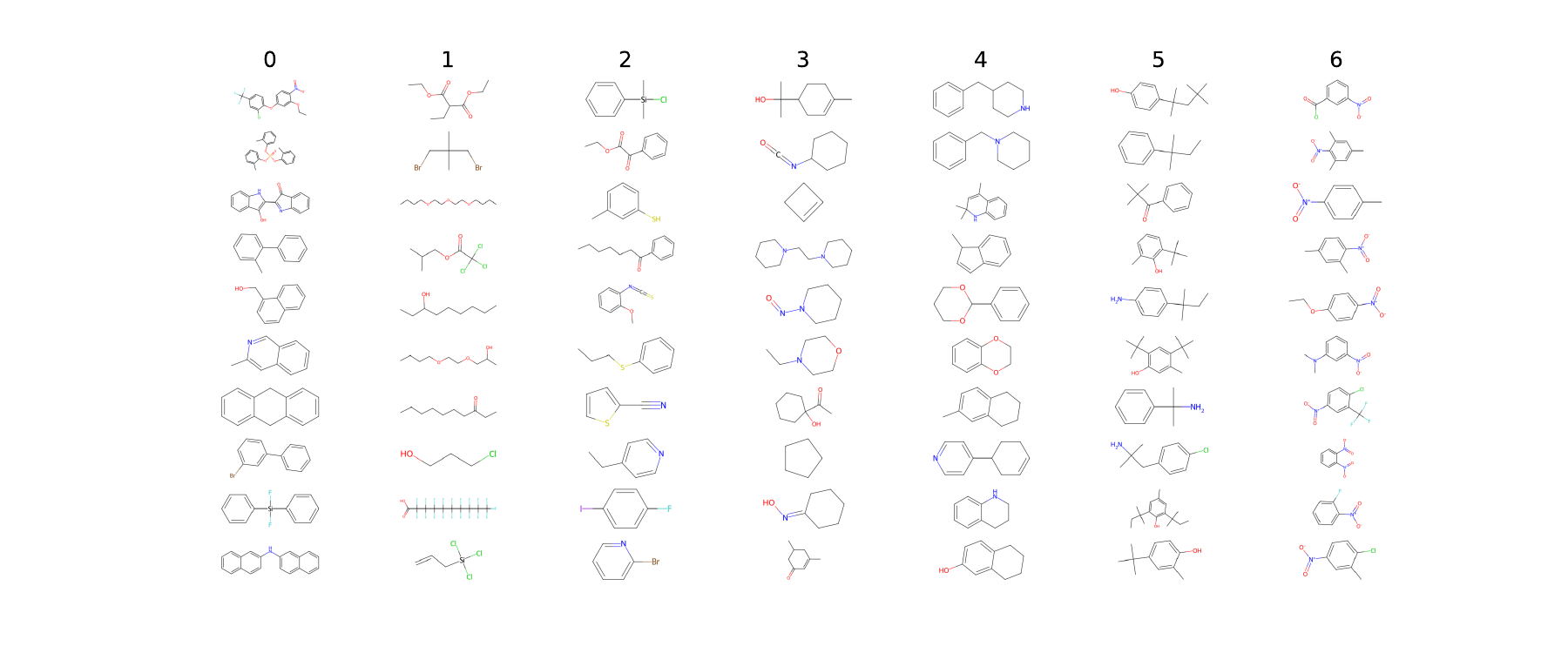}
        \caption{}
    \end{subfigure}
    \caption{BP Cluster analysis: (a) Average ring count per cluster for different ring types. (b) Representative molecular structures from each cluster}
\end{figure*}

\clearpage

\subsection{VP}

\begin{figure*}[!htb]
    \centering
    \begin{subfigure}[b]{0.7\linewidth}
        \centering
        \includegraphics[width=\linewidth]{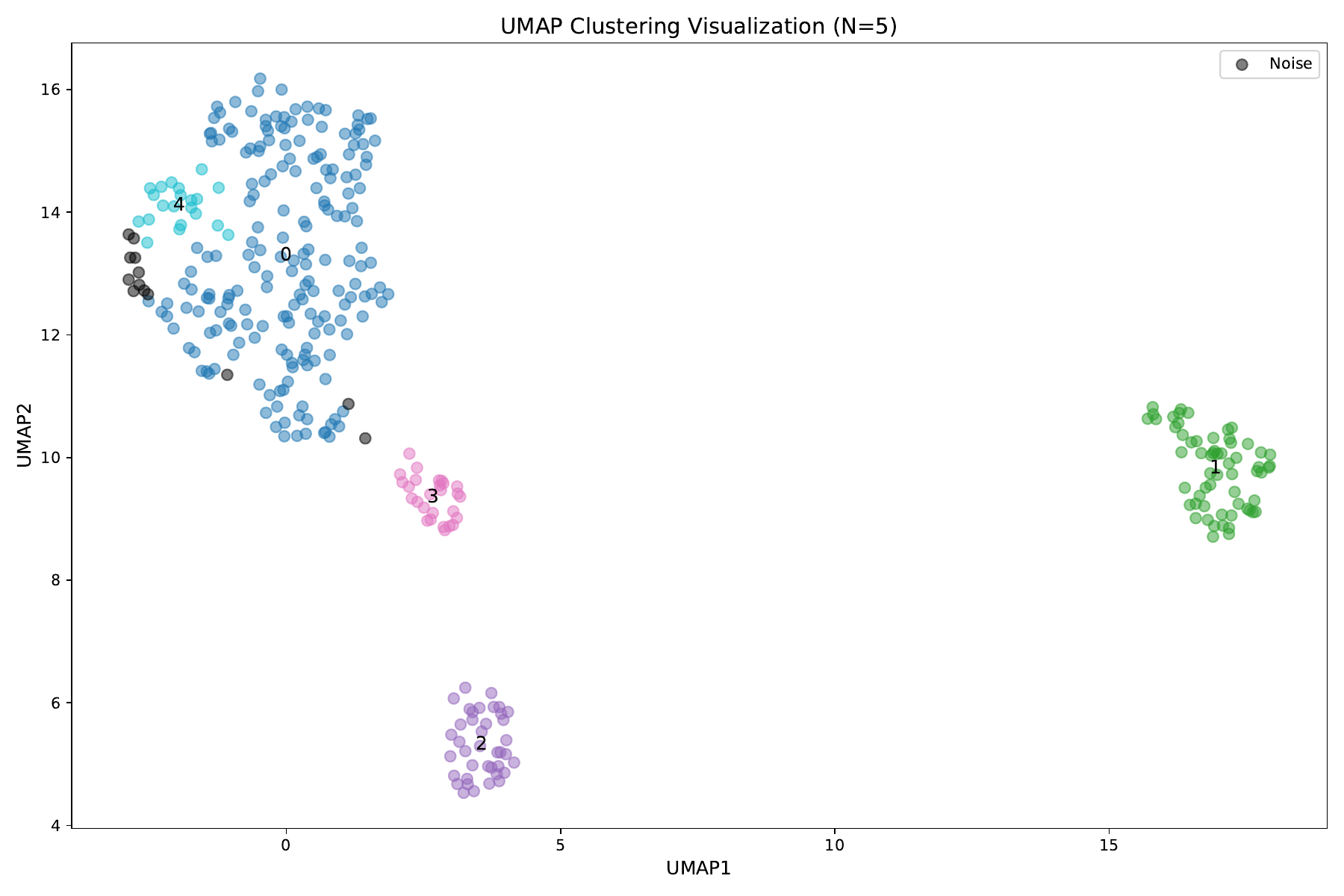}
        \caption{}
    \end{subfigure}
    \begin{subfigure}[b]{0.7\linewidth}
        \centering
        \includegraphics[width=\linewidth]{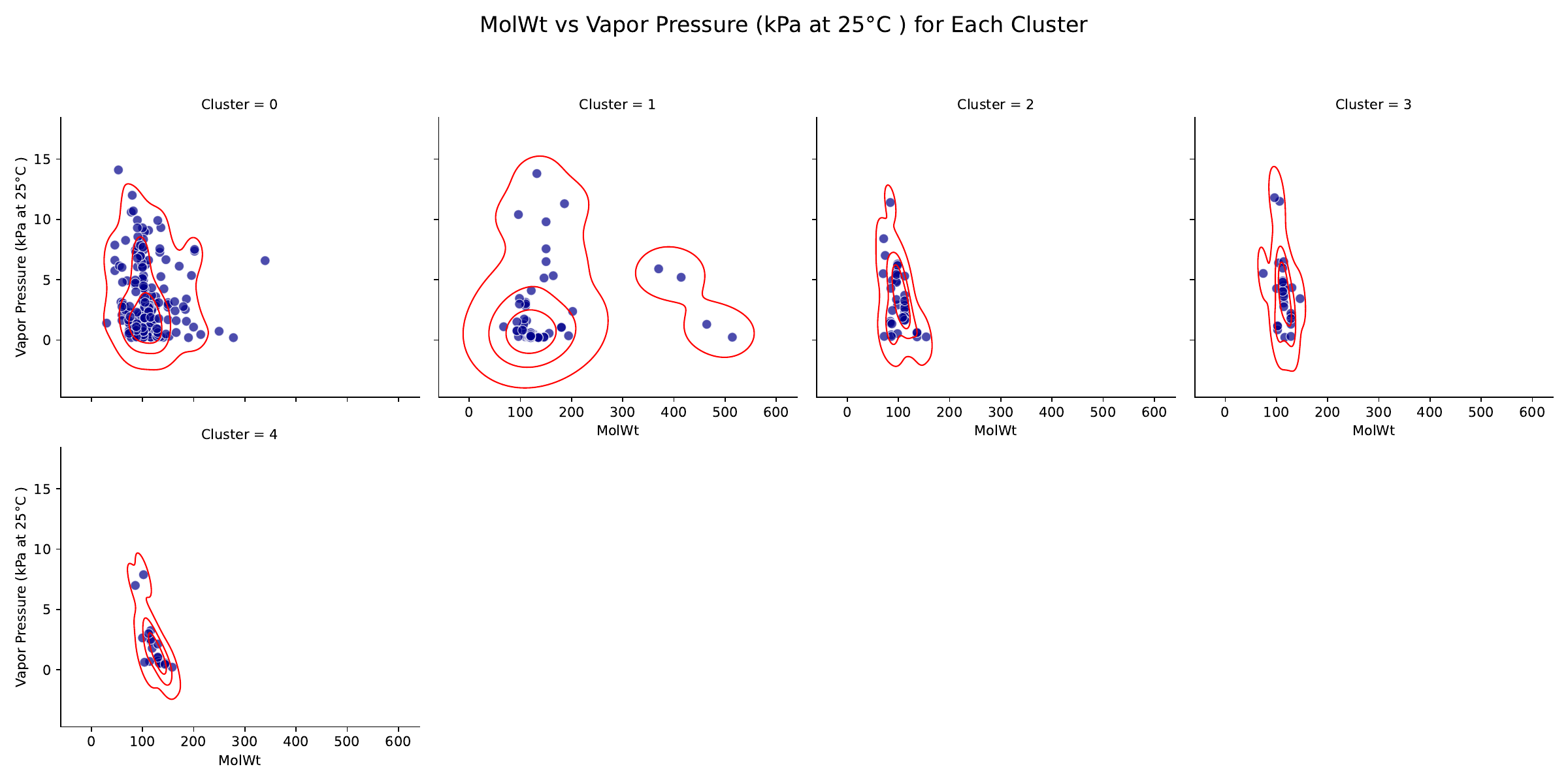}
        \caption{)}
    \end{subfigure}
    \caption{VP Cluster analysis: (a) UMAP representation of molecular embeddings, color-coded by identified clusters. (b) Relationship between molecular weight (MolWt, u) and VP (kPa at 25 $^\circ$C).}
\end{figure*}

\begin{figure*}[!htb]
    \centering
    \begin{subfigure}[b]{\linewidth}
        \centering
        \includegraphics[width=\linewidth]{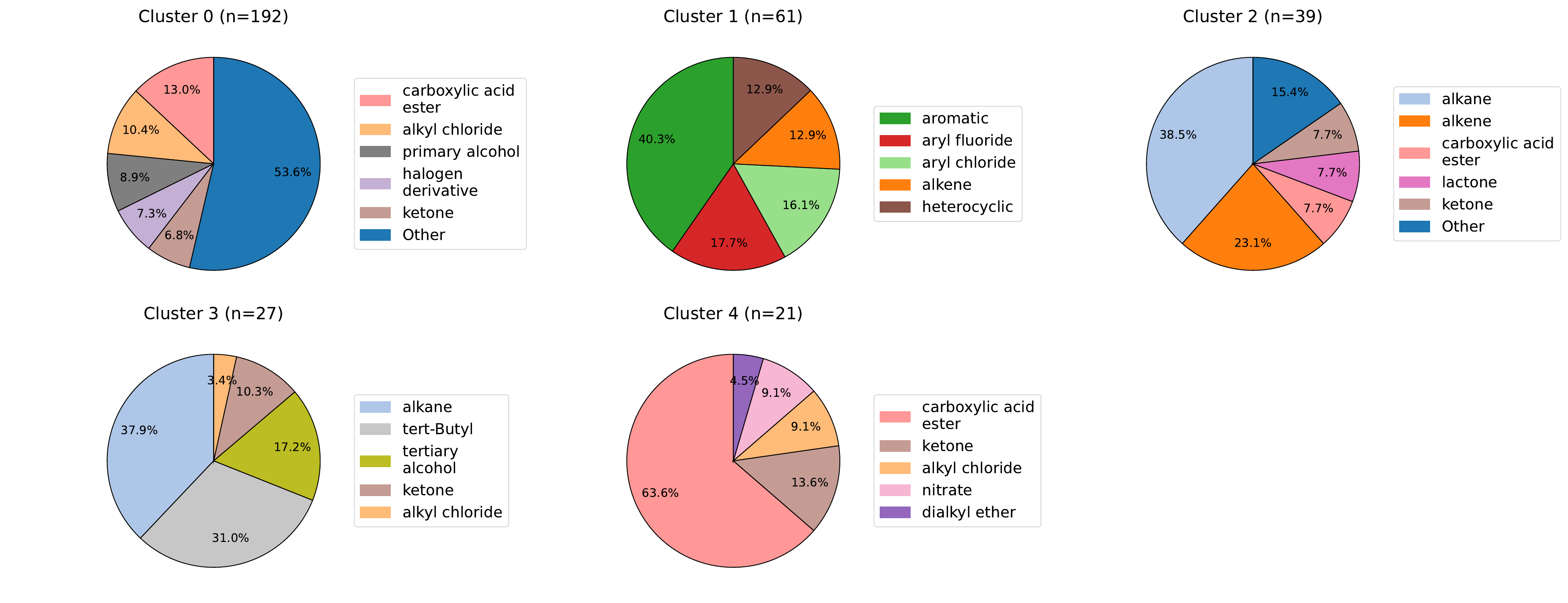}
        \caption{}
    \end{subfigure}
    \begin{subfigure}[b]{\linewidth}
        \centering
        \includegraphics[width=\linewidth]{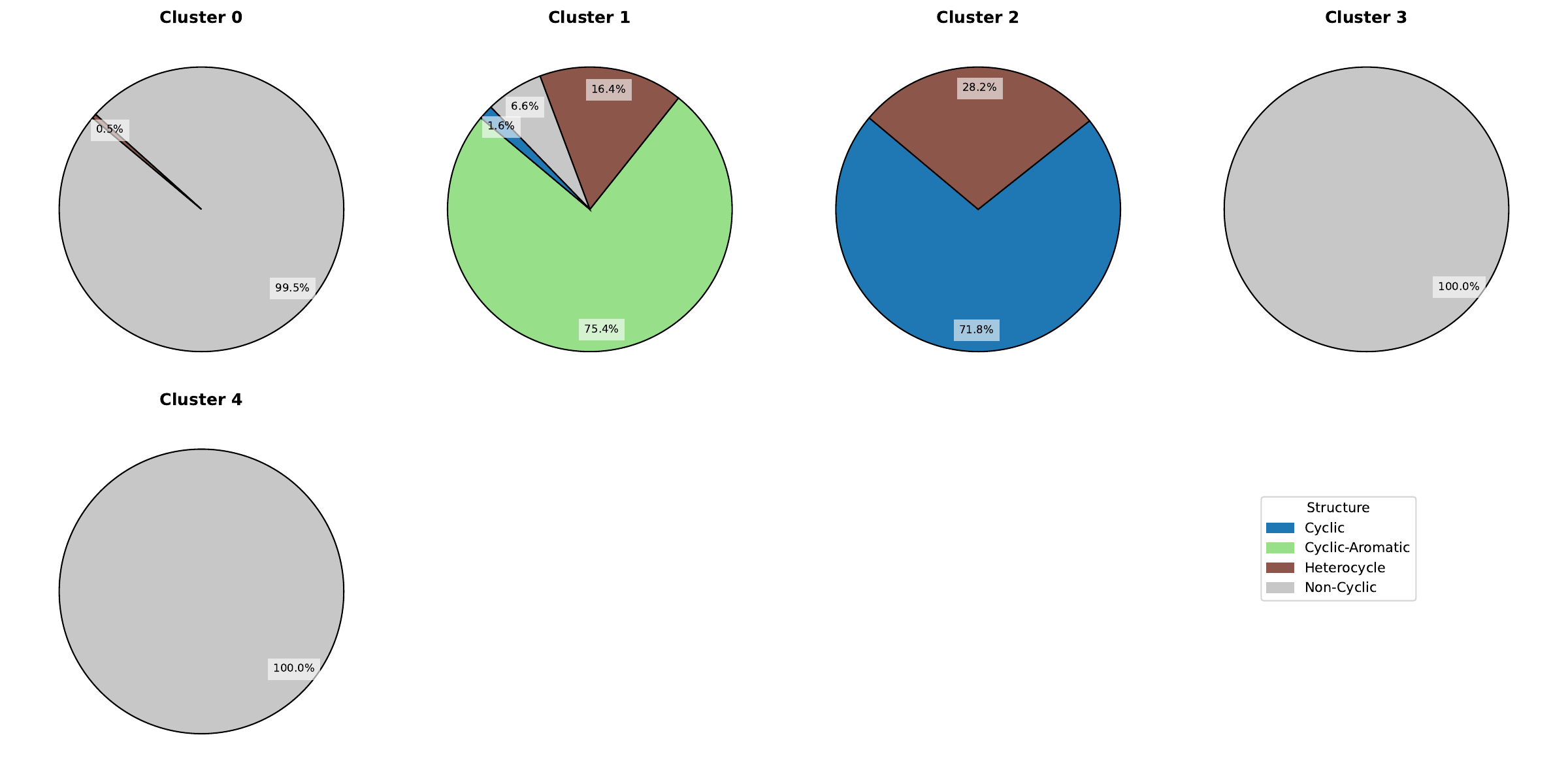}
        \caption{}
    \end{subfigure}
    \caption{VP Cluster analysis: (a) Functional group distribution across clusters. (b) Structural composition analysis shows the proportion of cyclic, aromatic, and non-cyclic compounds per cluster, with cyclic and highly conjugated systems favouring higher MPs.}
\end{figure*}

\begin{figure*}[!htb]
    \centering
     \begin{subfigure}[b]{\linewidth}
        \centering
        \includegraphics[width=\linewidth]{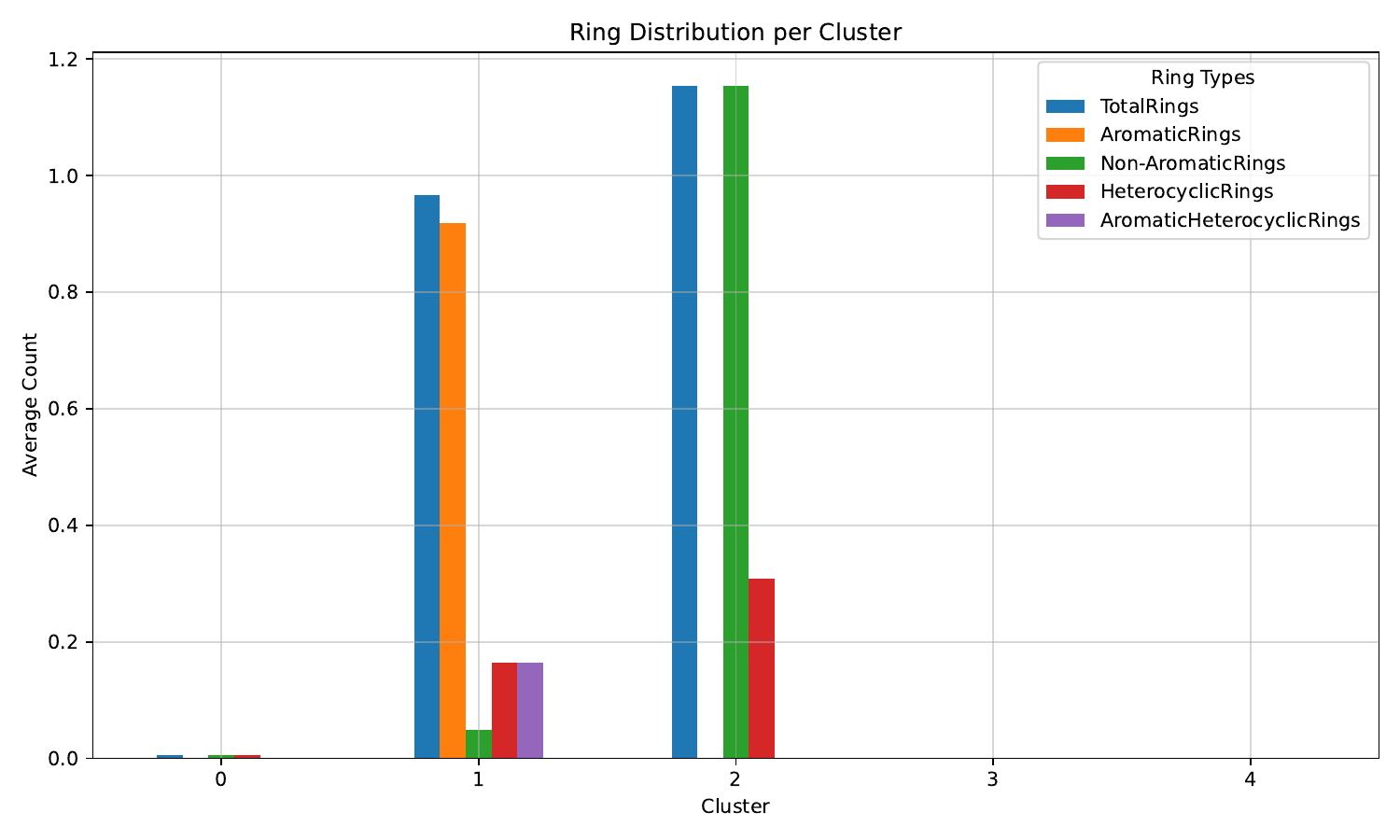}
        \caption{}
    \end{subfigure}
    \begin{subfigure}[b]{\linewidth}
        \centering
        \includegraphics[width=\linewidth]{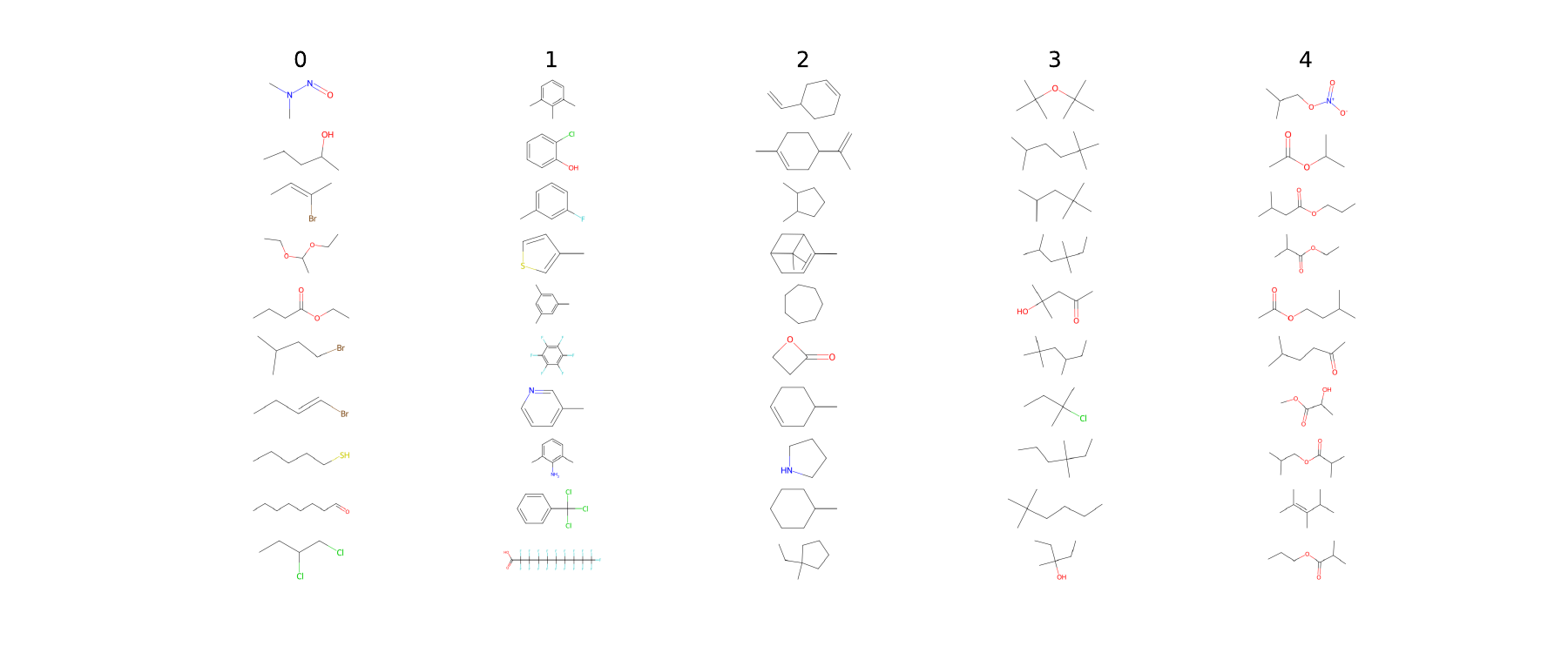}
        \caption{}
    \end{subfigure}
    \caption{VP Cluster analysis: (a) Average ring count per cluster for different ring types. (b) Representative molecular structures from each cluster}
\end{figure*}

\clearpage

\subsection{CT}

\begin{figure*}[!htb]
    \centering
    \begin{subfigure}[b]{0.7\linewidth}
        \centering
        \includegraphics[width=\linewidth]{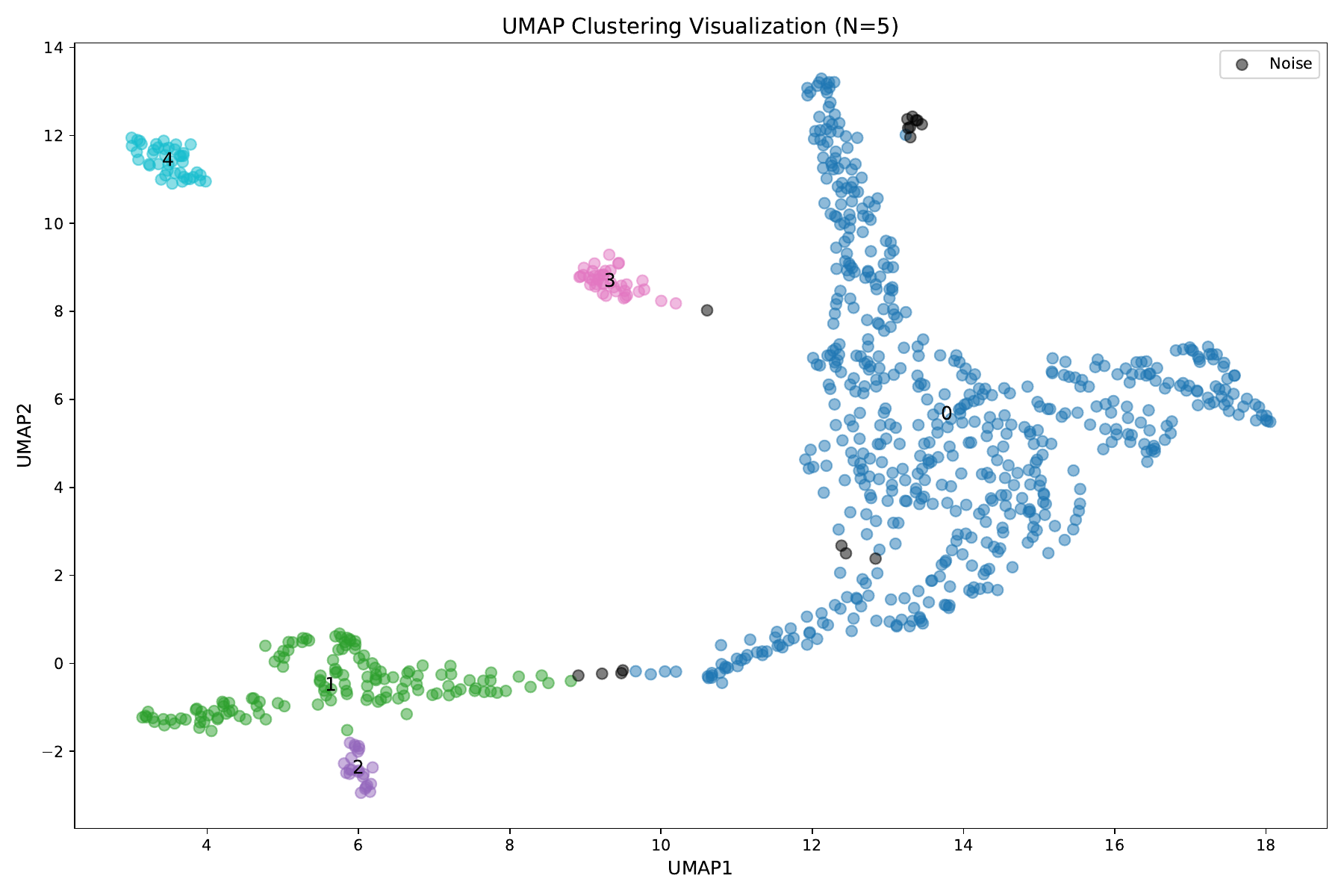}
        \caption{}
    \end{subfigure}
    \begin{subfigure}[b]{0.7\linewidth}
        \centering
        \includegraphics[width=\linewidth]{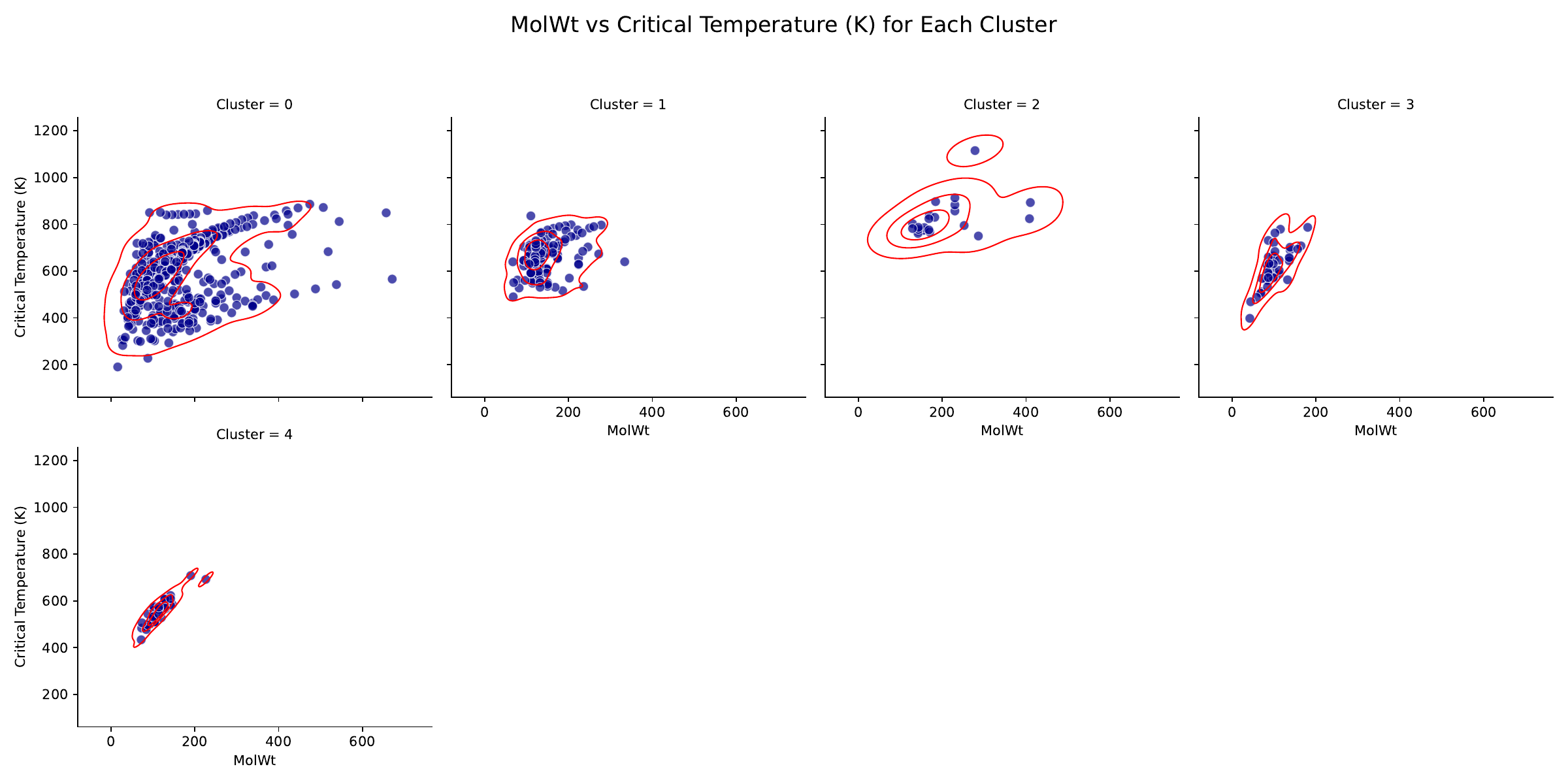}
        \caption{)}
    \end{subfigure}
    \caption{CT Cluster analysis: (a) UMAP representation of molecular embeddings, color-coded by identified clusters. (b) Relationship between molecular weight (MolWt, u) and CT (K).}
\end{figure*}

\begin{figure*}[!htb]
    \centering
    \begin{subfigure}[b]{\linewidth}
        \centering
        \includegraphics[width=\linewidth]{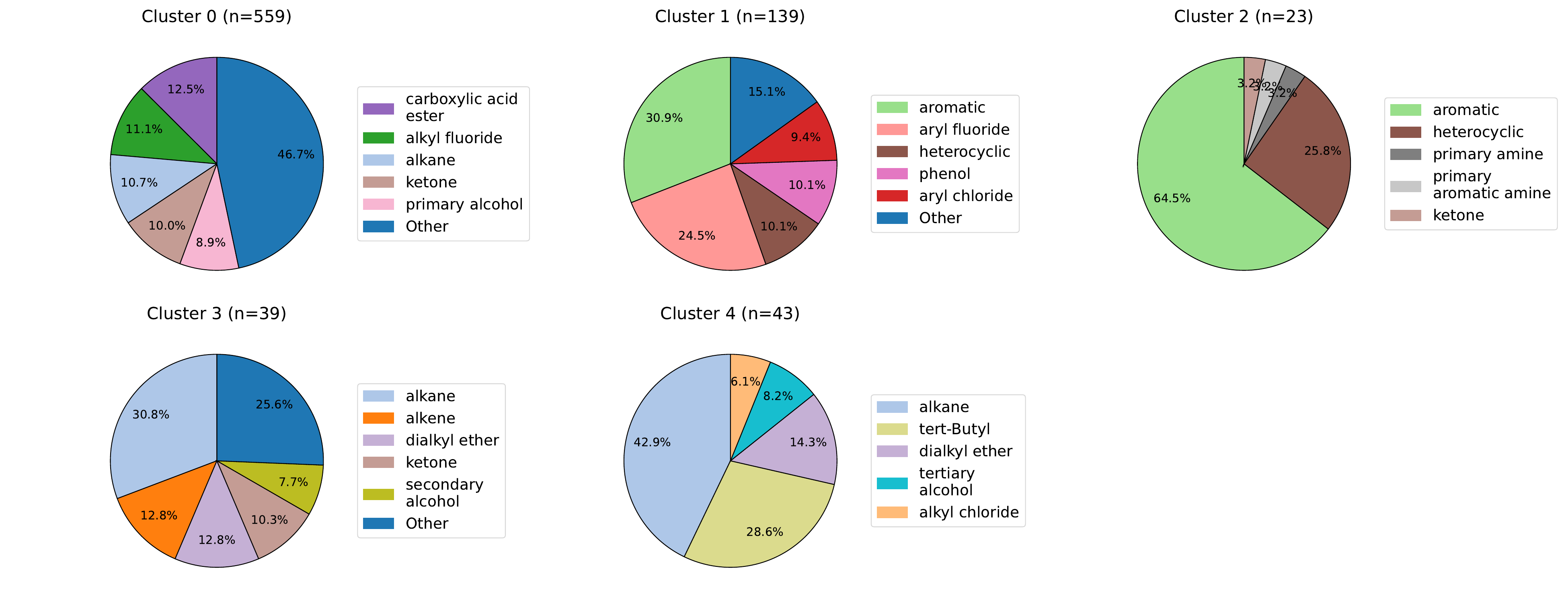}
        \caption{}
    \end{subfigure}
    \begin{subfigure}[b]{\linewidth}
        \centering
        \includegraphics[width=\linewidth]{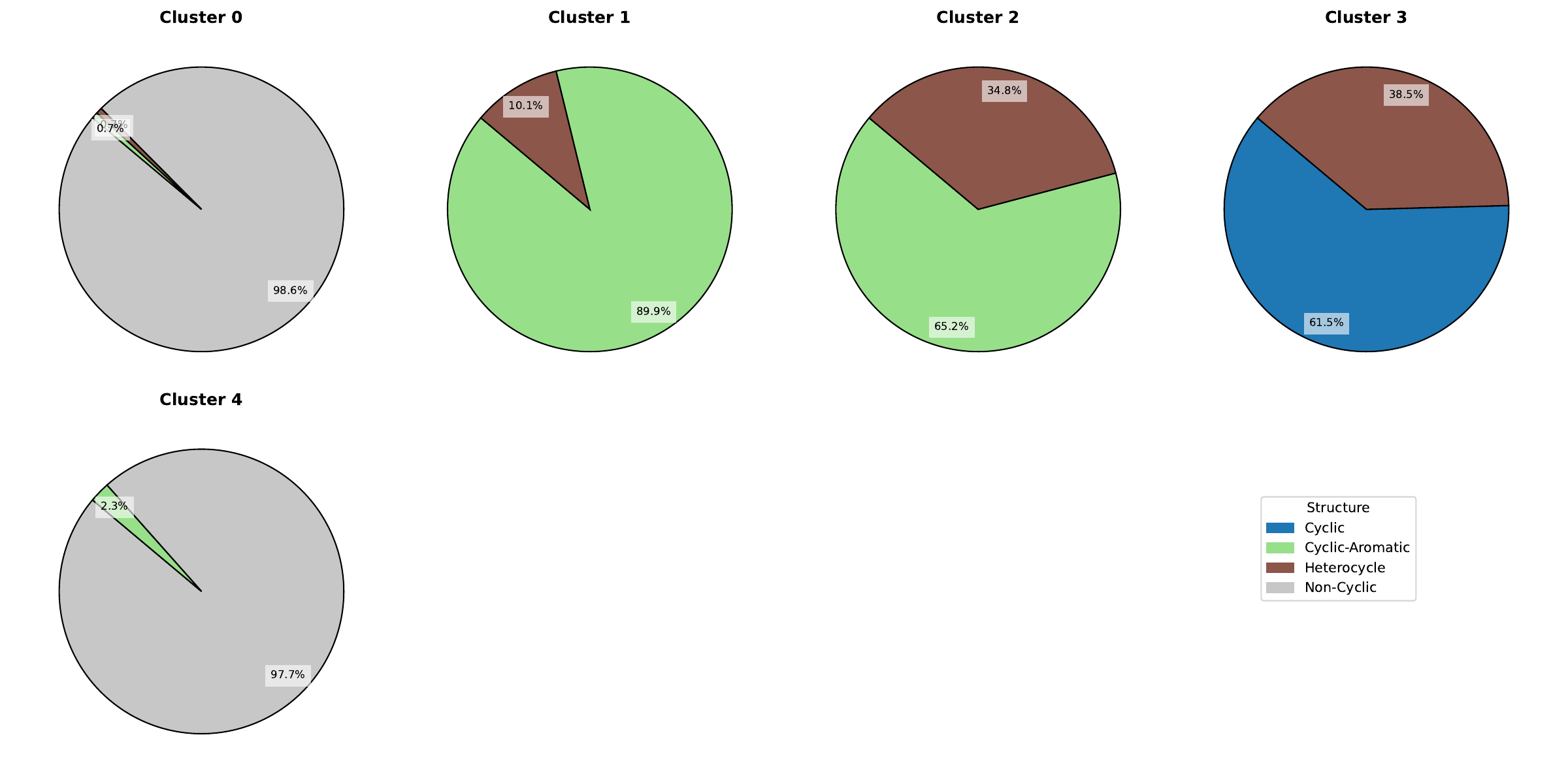}
        \caption{}
    \end{subfigure}
    \caption{CT Cluster analysis: (a) Functional group distribution across clusters. (b) Structural composition analysis shows the proportion of cyclic, aromatic, and non-cyclic compounds per cluster, with cyclic and highly conjugated systems favouring higher MPs.}
\end{figure*}

\begin{figure*}[!htb]
    \centering
     \begin{subfigure}[b]{\linewidth}
        \centering
        \includegraphics[width=\linewidth]{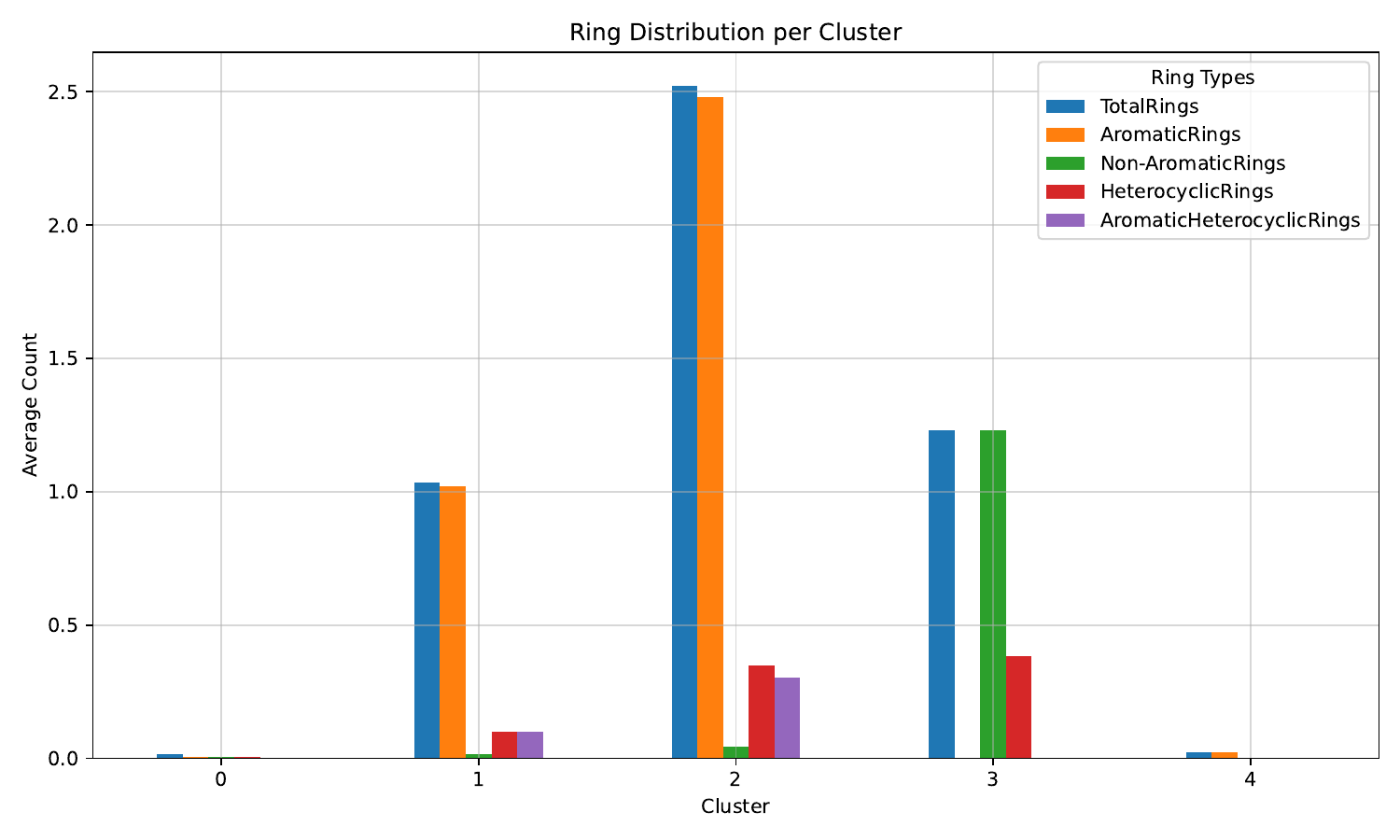}
        \caption{}
    \end{subfigure}
    \begin{subfigure}[b]{\linewidth}
        \centering
        \includegraphics[width=\linewidth]{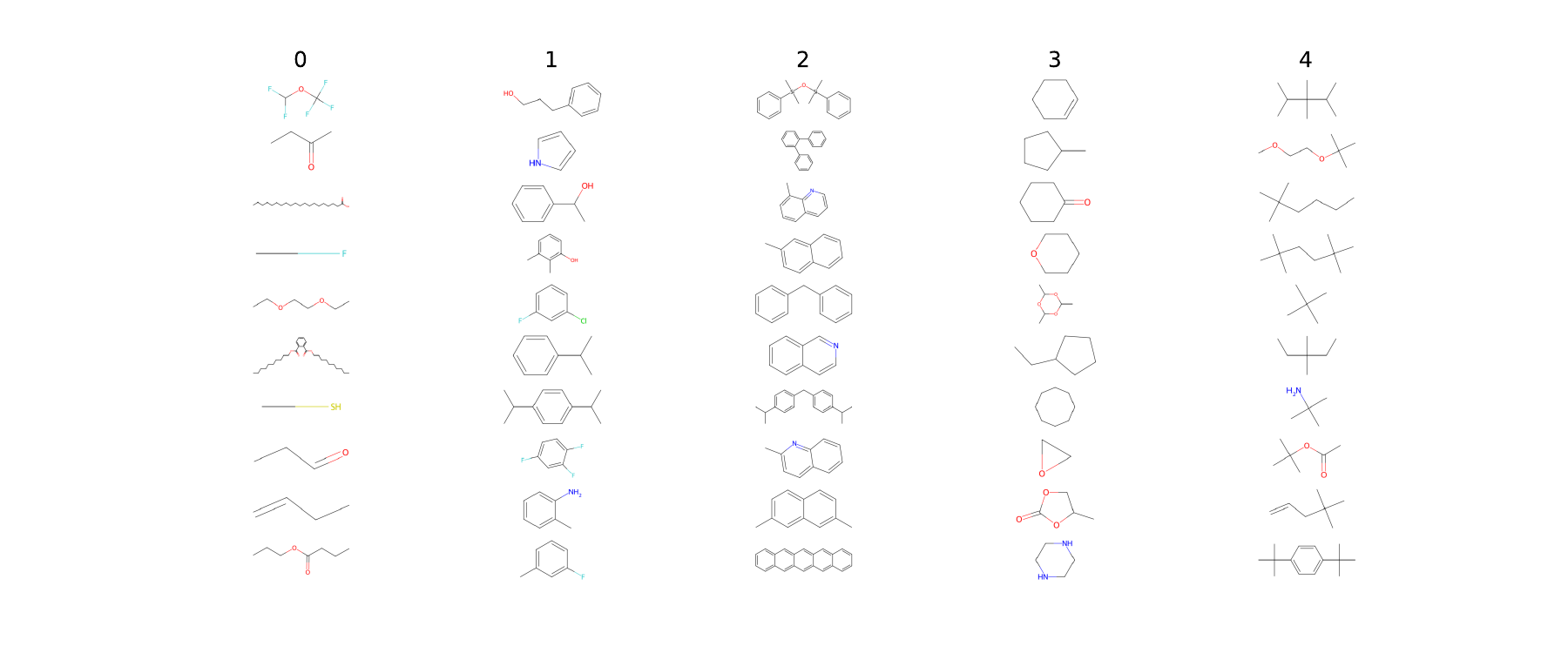}
        \caption{}
    \end{subfigure}
    \caption{CT Cluster analysis: (a) Average ring count per cluster for different ring types. (b) Representative molecular structures from each cluster}
\end{figure*}

\clearpage

\subsection{CP}

\begin{figure*}[!htb]
    \centering
    \begin{subfigure}[b]{0.7\linewidth}
        \centering
        \includegraphics[width=\linewidth]{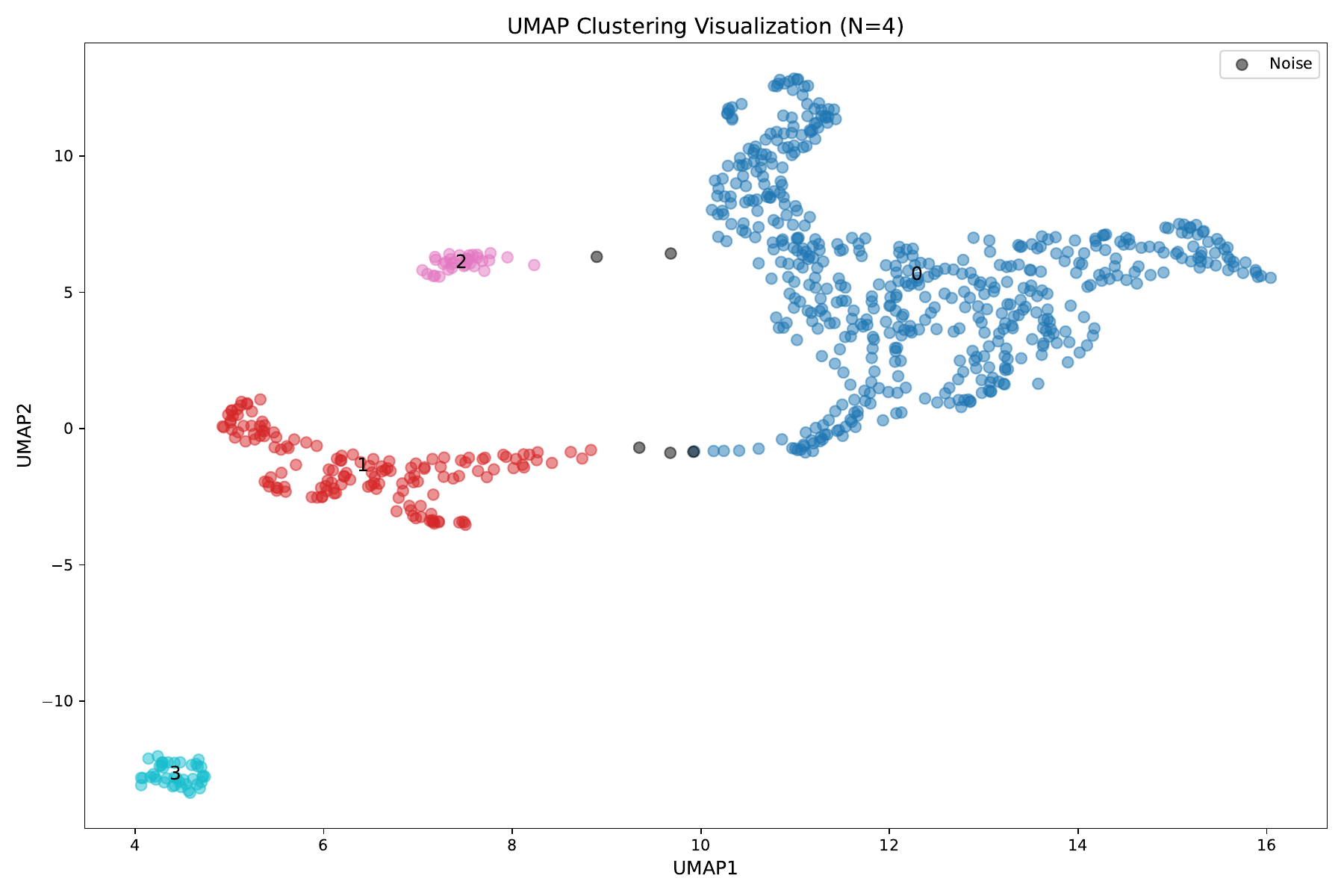}
        \caption{}
    \end{subfigure}
    \begin{subfigure}[b]{0.7\linewidth}
        \centering
        \includegraphics[width=\linewidth]{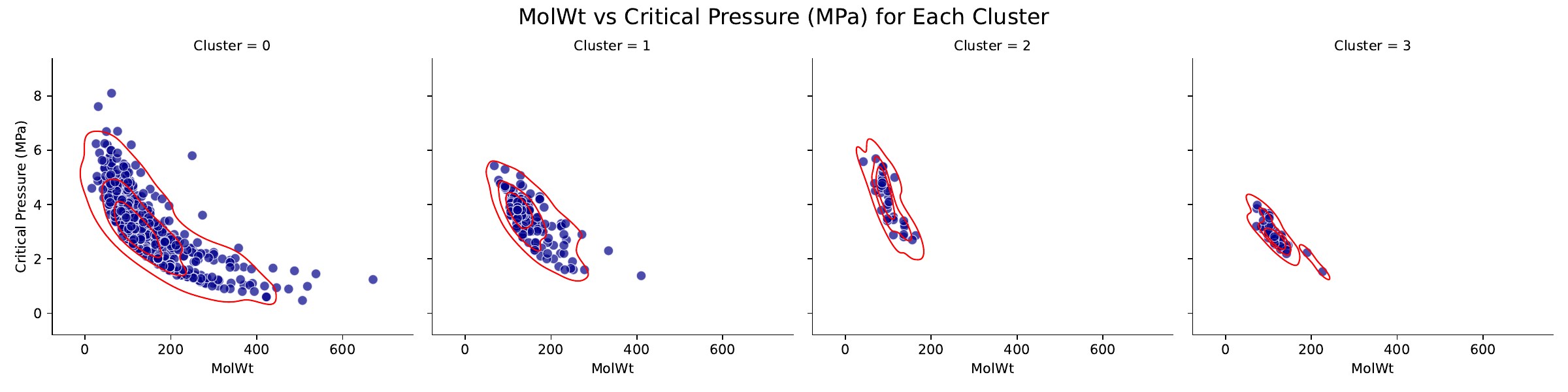}
        \caption{)}
    \end{subfigure}
    \caption{CP Cluster analysis: (a) UMAP representation of molecular embeddings, color-coded by identified clusters. (b) Relationship between molecular weight (MolWt, u) and CP (MPa).}
\end{figure*}

\begin{figure*}[!htb]
    \centering
    \begin{subfigure}[b]{\linewidth}
        \centering
        \includegraphics[width=\linewidth]{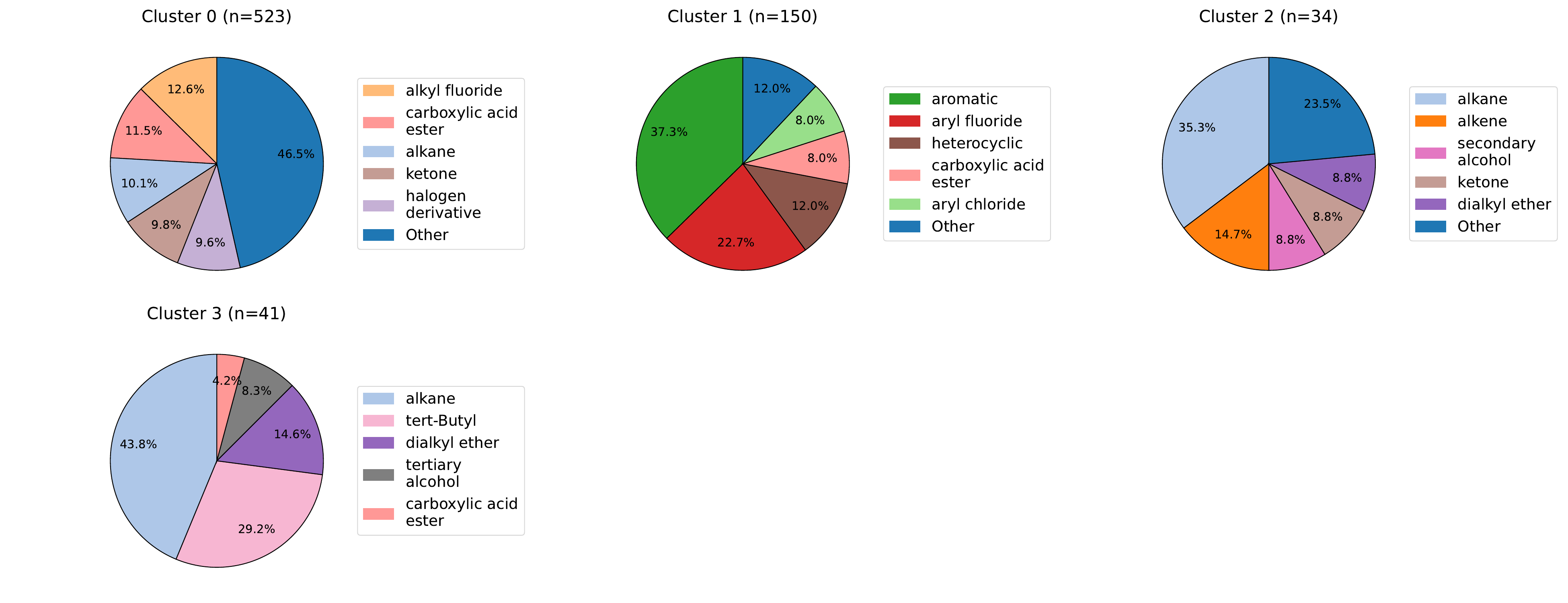}
        \caption{}
    \end{subfigure}
    \begin{subfigure}[b]{\linewidth}
        \centering
        \includegraphics[width=\linewidth]{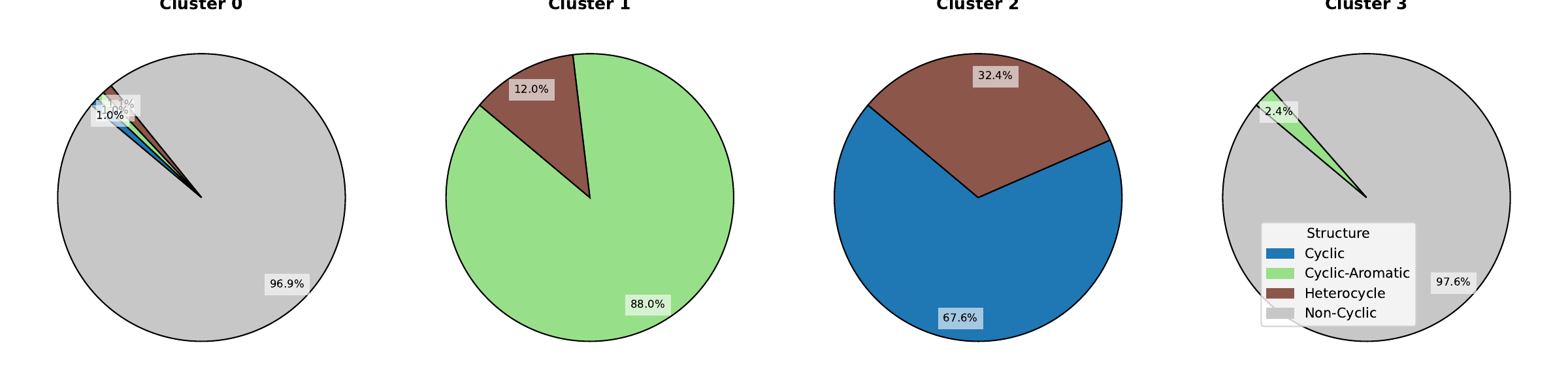}
        \caption{}
    \end{subfigure}
    \caption{CP Cluster analysis: (a) Functional group distribution across clusters. (b) Structural composition analysis shows the proportion of cyclic, aromatic, and non-cyclic compounds per cluster, with cyclic and highly conjugated systems favouring higher MPs.}
\end{figure*}

\begin{figure*}[!htb]
    \centering
     \begin{subfigure}[b]{\linewidth}
        \centering
        \includegraphics[width=\linewidth]{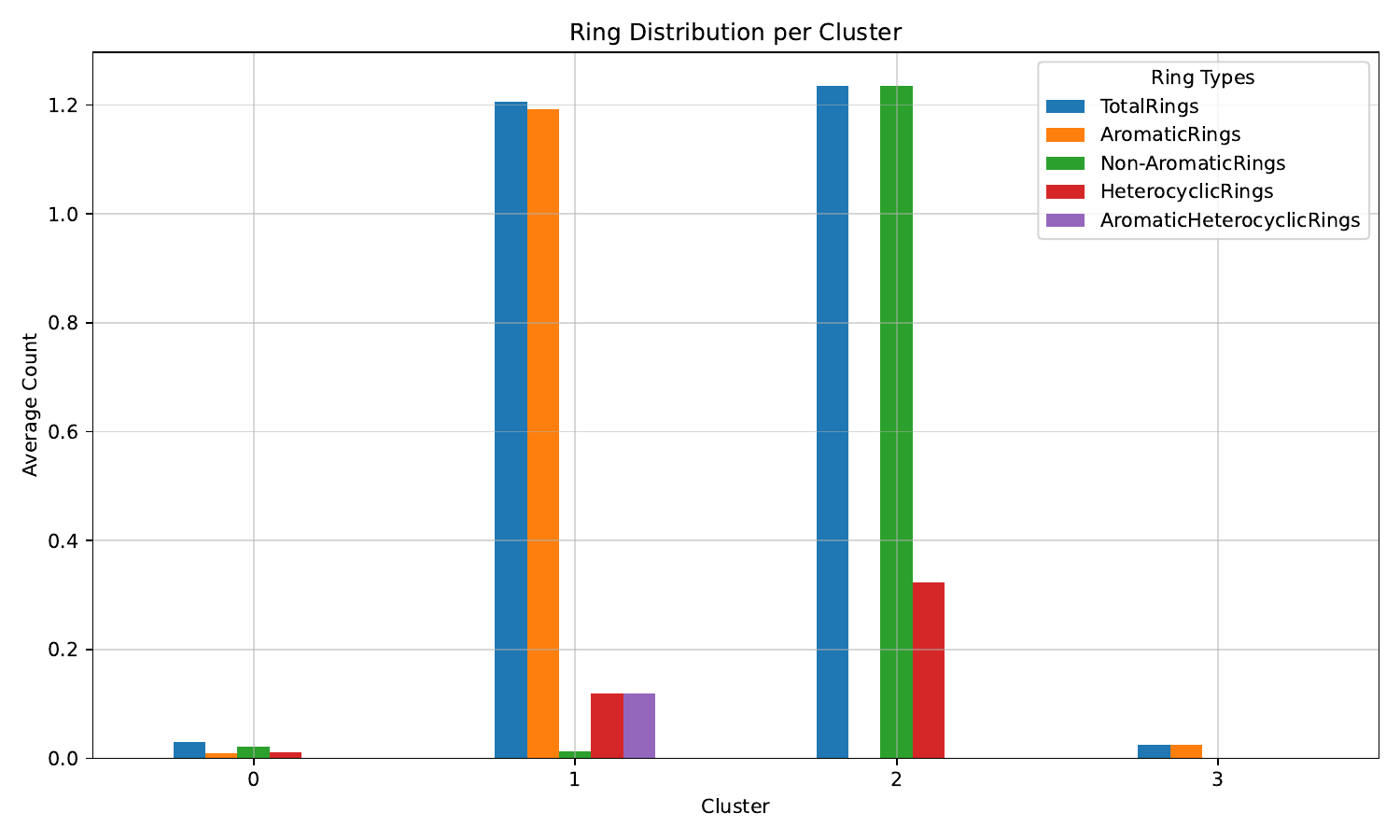}
        \caption{}
    \end{subfigure}
    \begin{subfigure}[b]{\linewidth}
        \centering
        \includegraphics[width=\linewidth]{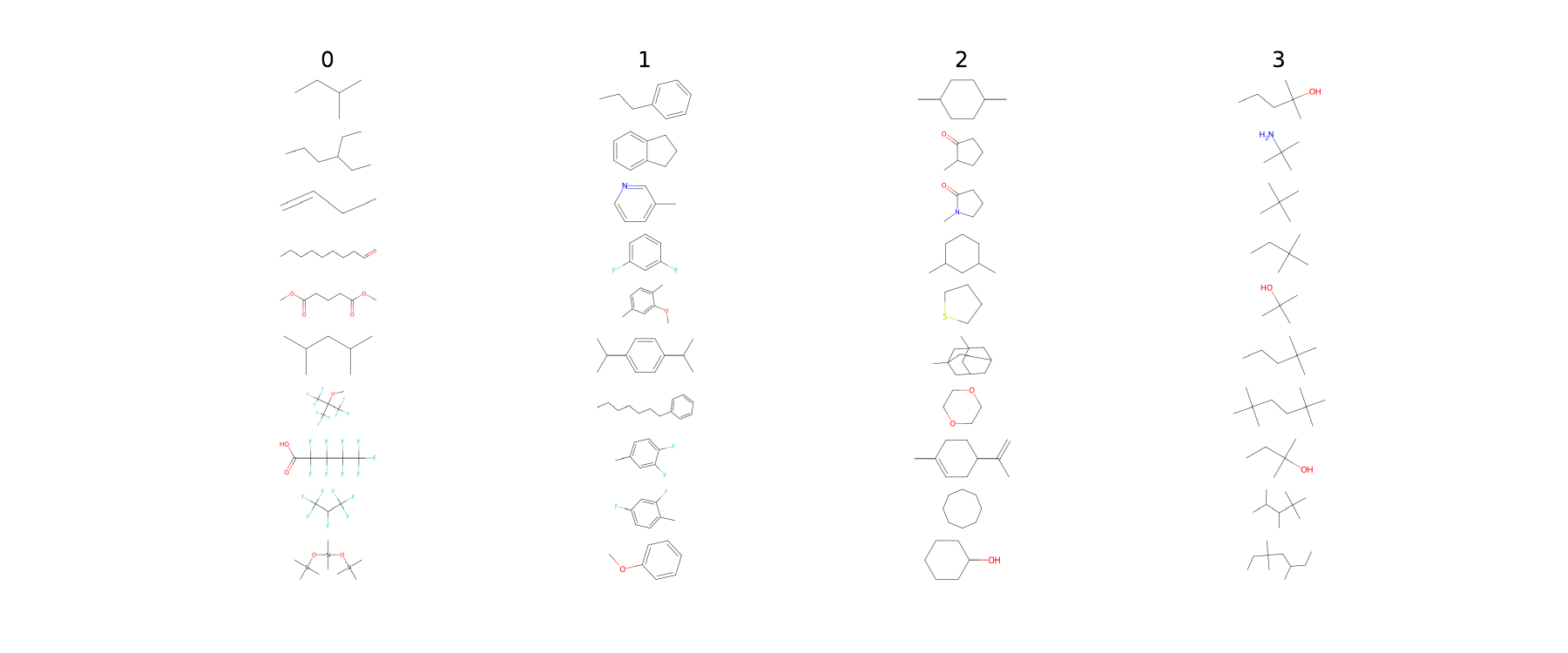}
        \caption{}
    \end{subfigure}
    \caption{CP Cluster analysis: (a) Average ring count per cluster for different ring types. (b) Representative molecular structures from each cluster}
\end{figure*}

\clearpage


\clearpage

\section{Installation Instructions}

To install ChemXploreML on your system, download the appropriate binary from the official GitHub releases page:

\url{https://github.com/aravindhnivas/ChemXploreML/releases}

\subsection*{Step 1: Choose the Correct Installer}

Under the latest release (e.g., \texttt{v4.3.4}), select the appropriate installer for your operating system:

\begin{itemize}
  \item \textbf{Windows (64-bit):} Use \texttt{ChemXploreML\_4.3.4\_x64-setup.exe}
  \item \textbf{macOS (Apple Silicon M1/M2/M3/M4):} \texttt{ChemXploreML\_4.3.4\_aarch64.dmg}
  \item \textbf{macOS (Intel-based):} \texttt{ChemXploreML\_4.3.4\_x64.dmg}
  \item \textbf{Linux (Ubuntu/Debian):} \texttt{ChemXploreML\_4.4.0\_amd64.deb}
  \item \textbf{Linux (Fedora/RHEL):} \texttt{ChemXploreML-4.4.0-1.x86\_64.rpm}
  \item \textbf{Linux (Generic):} \texttt{ChemXploreML\_4.4.0\_amd64.AppImage}
\end{itemize}

\subsection*{Step 2: Install the Application}

\textbf{Windows:} Double-click the \texttt{.exe} installer and follow the on-screen instructions.

\textbf{macOS:} Mount the \texttt{.dmg} file and drag the ChemXploreML app into your \texttt{/Applications} folder.

\textbf{Linux (Deb/RPM):} Install using:
\begin{itemize}
  \item Debian/Ubuntu: \texttt{sudo dpkg -i ChemXploreML\_4.4.0\_amd64.deb}
  \item Fedora/RHEL: \texttt{sudo rpm -i ChemXploreML-4.4.0-1.x86\_64.rpm}
\end{itemize}

\textbf{Linux (AppImage):}  

Make the file executable and run:
\begin{verbatim}
chmod +x ChemXploreML_4.4.0_amd64.AppImage
./ChemXploreML_4.4.0_amd64.AppImage
\end{verbatim}

\subsection*{Step 3: macOS Gatekeeper Warning (Important)}

macOS may prevent launching unsigned apps. To bypass this:

\begin{enumerate}
  \item Open Terminal.
  \item Run: \texttt{xattr -c /Applications/ChemXploreML.app}
  \item Right-click the app and select “Open”.
\end{enumerate}

A notarized version will be released in a future update.

\subsection*{Step 4: Launching and Using ChemXploreML}

Once installed, launch ChemXploreML from your system's app launcher. The application runs entirely offline and enables dataset loading, molecular embedding, dimensionality reduction, model training, prediction, and analysis—all via an interactive interface powered by a Python backend.

\clearpage
